\documentstyle[epsf,prd,aps,aps12]{revtex}

\newcommand{\be}{\begin{equation}}
\newcommand{\ee}{\end{equation}}
\newcommand{\bea}{\begin{eqnarray}}
\newcommand{\eea}{\end{eqnarray}}
\newcommand{\nn}{\nonumber}

\begin{document}
\draft \maketitle
\begin{center}

\hfill\break

\hfill\break

{\Large\textbf{TRABAJO DE DIPLOMA}}

\bigskip

\bigskip

\bigskip

\bigskip

\bigskip

{\Large\textbf{ENERG\'{I}AS DE VAC\'{I}O PARA EL PROBLEMA DE LANDAU
RELATIVISTA}}

\bigskip

\bigskip

\bigskip

\bigskip

\bigskip

\bigskip

{\Large\textbf{Diego H. Correa}}

\bigskip

\bigskip

\bigskip

\bigskip

\bigskip

{\large\textbf{Directora: E. M. Santangelo}}

\bigskip

\bigskip

{\large\textbf{Departamento de F\'{\i}sica}}
\\ {\large\textbf{Facultad de Ciencias Exactas}}\\
 {\large\textbf{UNLP}}

\bigskip

{\large\textbf{agosto de 2000}}
\end{center}

\newpage

\section{Introducci\'{o}n}
\label{sec-int}

El m\'{e}todo can\'{o}nico de cuantizaci\'{o}n permite establecer una
correspondencia entre las cantidades cl\'{a}sicas observables y los
operadores cu\'{a}nticos. El hecho de que al emplearlo para cuantizar
campos aparezcan energ\'{\i}as de vac\'{\i}o divergentes es una consecuencia
de la ambig\"{u}edad en la elecci\'{o}n del orden usado para escribir los
t\'{e}rminos del hamiltoniano, que son productos de operadores que no
conmutan entre s\'{\i}. Los hamiltonianos escritos con diferentes
\'{o}rdenes difieren en t\'{e}rminos proporcionales al operador identidad
\cite{itzykson}. Restar estos t\'{e}rminos para pasar de un
ordenamiento a otro es equivalente a trasladar el nivel cero de
energ\'{\i}a. Ya que solamente las diferencias de energ\'{\i}as entre
estados son observables, la elecci\'{o}n del cero de energ\'{\i}a no afecta
la din\'{a}mica de los campos. As\'{\i}, en la teor\'{\i}a de campos libres, la
ambig\"{u}edad se resuelve con la prescripci\'{o}n de orden normal que
conduce a un valor medio de vac\'{\i}o nulo para el operador
hamiltoniano.

En 1948, H. B. G. Casimir mostr\'{o} que la presencia de dos placas
conductoras neutras conduce a un valor medio de vac\'{\i}o no nulo para
el operador hamiltoniano de un campo
electromagn\'{e}tico\cite{casimir}.

Por extensi\'{o}n, se conoce como efecto Casimir a la variaci\'{o}n que sufren los
valores medios de vac\'{\i}o de observables de un campo al impon\'{e}rsele condiciones
de contorno externas o al interactuar con un campo externo generado por determinada
configuraci\'{o}n de fuentes. La diferencia entre las energ\'{\i}as de vac\'{\i}o de
estas dos configuraciones se define como Energ\'{\i}a de Casimir. As\'{\i}, se tiene

\begin{equation}
E_{Cas}=E_{0}-E_{0}^{l} .\label{Ecas}\end{equation} donde $E_{0}^{l}$ es la
energ\'{\i}a de vac\'{\i}o del campo libre y $E_{0}$ es la energ\'{\i}a de vac\'{\i}o
del campo distorsionada por condiciones de contorno o la interacci\'{o}n con un campo
externo.

Los m\'{e}todos de evaluaci\'{o}n de energ\'{\i}as de vac\'{\i}o pueden reunirse en
dos categor\'{\i}as \cite{Plunien}: el m\'{e}todo de suma directa de modos,
aplicable toda vez que el espectro cl\'{a}sico sea expl\'{\i}citamente
determinable, y el m\'{e}todo local, que requiere el conocimiento de
la funci\'{o}n de Green del problema. En este trabajo, usaremos el
m\'{e}todo de suma de modos. Seg\'{u}n \'{e}ste, la energ\'{\i}a de vac\'{\i}o se
obtiene de sumar los autovalores de energ\'{\i}a de los infinitos modos
del problema particular. Para el caso de un campo de Dirac como el
que trataremos, la energ\'{\i}a de vac\'{\i}o resulta \cite{libro-cas}

\begin{equation}\label{E0}
  E_{0}=-\frac{1}{2}\sum_{k}g_{k}|E_{k}|,
\end{equation}
donde $E_{k}$ son los autovalores del espectro cl\'{a}sico de energ\'{\i}as,
correspondientes a autofunciones de cuadrado integrable, y $g_{k}$ la degeneraci\'{o}n
de cada uno de ellos.

\medskip

La suma en (\ref{E0}) resulta, en general, divergente. A fin de dar una
interpretaci\'{o}n a las divergencias y poder calcular (\ref{Ecas}) debe recurrirse a
alguna t\'{e}cnica de regularizaci\'{o}n. En este trabajo, usaremos el m\'{e}todo de
regularizaci\'{o}n de la funci\'{o}n zeta \cite{Dowker:1976tf,zeta}.

La funci\'{o}n zeta de un operador $\hat{A}$ es definida, cuando son conocidos sus
autovalores $\lambda_{n}$, como

\begin{equation}
\zeta_{A}=\sum_{n}\lambda_{n}^{-s}.
\end{equation}

Esta suma resulta convergente s\'{o}lo para $Re(s)$ mayor que cierta cantidad, que
depende de la dimensi\'{o}n del espacio y del orden del operador \cite{libro-gilkey}.
Sin embargo, la funci\'{o}n puede extenderse anal\'{\i}ticamente a todo el plano
complejo, como una funci\'{o}n meromorfa con polos en un conjunto discreto de valores
de $s$.

En nuestro caso, definiremos la suma (\ref{E0}) como

\begin{equation}\label{E0-s}
  E_{0}=\left(-\frac{\mu^{s+1}}{2}\sum_{k}g_{k}|E_{k}|^{-s}\right)_{s=-1},
\end{equation}
donde $\mu$ es un par\'{a}metro de escala con unidades de energ\'{\i}a.

Es decir: con el objeto de evaluar las energ\'{\i}as de vac\'{\i}o que aparecen en la
definici\'{o}n (\ref{Ecas}) de la energ\'{\i}a de Casimir, realizaremos la
extensi\'{o}n anal\'{\i}tica de la funci\'{o}n zeta correspondiente al hamiltoniano
cl\'{a}sico, evalu\'{a}ndola posteriormente en el valor $s=-1$ de su extensi\'{o}n
anal\'{\i}tica. Al hacerlo, pueden ocurrir dos cosas: que el procedimiento arroje un
resultado finito o bien que la extensi\'{o}n presente un polo en el valor de $s$ de
inter\'{e}s. Cuando esto \'{u}ltimo ocurre, debe realizarse una renormalizaci\'{o}n de
los par\'{a}metros en el hamiltoniano cl\'{a}sico, a fin de dar una interpretaci\'{o}n
f\'{\i}sica al resultado. En lo que sigue, veremos ejemplos de ambas situaciones.

El contenido del presente trabajo es el siguiente:

En la secci\'{o}n II, determinamos el operador hamiltoniano cl\'{a}sico para fermiones
en un campo magn\'{e}tico uniforme en la direcci\'{o}n del eje $z$, de acuerdo al gauge
elegido para el potencial electromagn\'{e}tico y la representaci\'{o}n elegida de la
matrices de Dirac.

En la secci\'{o}n III, obtenemos los autovalores del espectro cl\'{a}sico de
energ\'{\i}a, en caso en el que los fermiones est\'{a}n confinados en un plano
perpendicular al campo magn\'{e}tico, hallando las autofunciones de cuadrado integrable
del hamiltoniano cl\'{a}sico con impulso $p_{z}=0$.

En la secci\'{o}n IV, obtenemos las energ\'{\i}as de vac\'{\i}o y de Casimir del
problema en $2+1$ dimensiones, mediante la t\'{e}cnica de regularizaci\'{o}n de la
funci\'{o}n zeta.

En la secci\'{o}n V, determinamos el espectro cl\'{a}sico de energ\'{\i}a cuando todo
el espacio tridimensional es accesible para los fermiones. Lo hacemos aplicando una
transformaci\'{o}n de Lorentz sobre las autofunciones obtenidas en la secci\'{o}n III.

En la secci\'{o}n VI, obtenemos las energ\'{\i}as de vac\'{\i}o y de Casimir del
problema en \mbox{$3+1$} dimensiones, por medio de una regularizaci\'{o}n zeta y una
renormalizaci\'{o}n de la carga el\'{e}ctrica.

En la secci\'{o}n VII, consideraremos la variaci\'{o}n en la energ\'{\i}a de vac\'{\i}o
por efecto conjunto del campo magn\'{e}tico y condiciones de contorno en la
direcci\'{o}n del campo. Aqu\'{\i} nuevamente relizamos una regularizaci\'{o}n zeta y
una renormalizaci\'{o}n de la carga el\'{e}ctrica.

\newpage

\section{Generalidades}
\label{Generalidades}

Estudiaremos el problema de un campo de Dirac en interacci\'{o}n
con un campo electromagn\'{e}tico, caracterizado por su potencial
$A_{\mu}(x)$. La ecuaci\'{o}n de Dirac en el espacio de Minkowski
de $3+1$ dimensiones y en unidades naturales, $\hbar=c=1$, se
escribe entonces

\begin{equation}
(i\gamma^{\mu}\partial_{\mu}-e\gamma^{\mu}A_{\mu}-m)\Psi=0.
\end{equation}

Trataremos el caso de un campo magn\'{e}tico uniforme $\vec{B}$ en la direcci\'{o}n del
eje $z$. Sabemos que la elecci\'{o}n de un potencial no es \'{u}nica y que el operador
hamiltoniano depende de esta elecci\'{o}n. Eligiendo el potencial electromagn\'{e}tico
$A_{\mu}=(0,\frac{B}{2}y,-\frac{B}{2}x,0)$ los operadores hamiltoniano y componente $z$
del momento angular total conmutan y, por lo tanto, tendremos autofunciones
simult\'{a}neas de ambos operadores.

\medskip

Utilizaremos la siguiente representaci\'{o}n de las matrices de
Dirac.

\begin{equation}
\gamma^0 = \left(
\begin{array}{cc}
  \sigma_3 & 0 \\
  0 & \sigma_3
\end{array}
\right) \quad \gamma^1 = \left(
\begin{array}{cc}
  -i \sigma_2 & 0 \\
  0 & -i \sigma_2
\end{array}
\right) \quad \gamma^2 = \left(
\begin{array}{cc}
  i\sigma_1 & 0 \\
  0 & -i\sigma_1
\end{array}
\right) \quad \gamma^3 = \left(
\begin{array}{cc}
  0 & i \sigma_1 \\
  i \sigma_1 & 0
\end{array}
\right) ,
\end{equation}
donde $\sigma_{i}$ son las matrices de Pauli.

\medskip

A fin de determinar el espectro cl\'{a}sico de energ\'{\i}as,
factorizamos la dependencia temporal del espinor de Dirac y lo
escribimos en t\'{e}rminos de espinores de dos componentes

\begin{equation}
\Psi(x)=\psi_{E}(\vec{x})e^{-iEt}=\left(\begin{array}{c}
  \phi(\vec{x}) \\
  \chi(\vec{x})
\end{array}\right)e^{-iEt}.
\end{equation}

Obtenemos entonces

\be\label{H4} \hat{H}\left(\begin{array}{c}
  \phi(\vec{x}) \\
  \chi(\vec{x})
\end{array}\right)=E\left(\begin{array}{c}
  \phi(\vec{x}) \\
  \chi(\vec{x})
\end{array}\right) ,
\ee donde el operador hamiltoniano est\'{a} dado, en el caso de campo el\'{e}ctrico
nulo, por

\begin{equation}
H = \gamma^{0}\vec{\gamma}(-i\vec{\nabla}-e\vec{A})+m\gamma^{0}
=\left(
\begin{array}{cc}
 H_{+} & i\sigma_{2}\partial_{z} \\
  i\sigma_{2}\partial_{z} & H_{-}
\end{array}
\right) ,
\end{equation}
con

\begin{equation}
H_{\pm} = \left(
\begin{array}{cc}
 m & i\partial_{x}\pm\partial_{y}-\frac{eB}{2}(y\pm ix)
\\
  i\partial_{x}\mp\partial_{y}-\frac{eB}{2}(y\mp ix)
& -m
\end{array}
\right)
\end{equation}
o, en coordenadas cil\'{\i}ndricas

\begin{equation}
H_{\pm} = \left(
\begin{array}{cc}
 m & ie^{\mp i\theta}
\left(\partial_{r}\mp\frac{i}{r}\partial_{\theta}\mp\frac{eB}{2}r\right)
\\
  ie^{\pm i\theta}
\left(\partial_{r}\pm\frac{i}{r}\partial_{\theta}\pm\frac{eB}{2}r\right)
& -m
\end{array}
\right) .\label{H2}\end{equation}

\bigskip

En lo siguiente estudiaremos, en primer lugar, el problema de un campo de Dirac
confinado en un plano perpendicular al campo magn\'{e}tico. A continuaci\'{o}n
analizaremos el caso en el que todo el espacio es accesible a los fermiones y,
finalmente, estudiaremos el efecto de condiciones de contorno en la direcci\'{o}n del
campo.

\newpage

\section{Espectro en $2+1$ dimensiones}
\label{Esp-2+1}

Si los electrones est\'{a}n confinados en un plano perpendicular
al campo magn\'{e}tico, entonces las autofunciones resultan
independientes de $z$

\begin{equation}
\Psi(x)=\left(\begin{array}{c}
  \phi(r,\theta) \\
  \chi(r,\theta)
\end{array}\right)e^{-iEt} ,
\end{equation}
y el operador hamiltoniano resulta diagonal en bloques

\begin{equation}
H = \left(
\begin{array}{cc}
 H_{+} & 0 \\
  0 & H_{-}
\end{array}
\right) ,
\end{equation}
con $H_{\pm}$ igual que en la ecuaci\'{o}n (\ref{H2}).

\medskip

Cada uno de estos bloques corresponde a una de las representaciones no equivalentes de
las matrices de Dirac en $2+1$ dimensiones \cite{Flekkoy}.

\medskip

Estudiaremos, en primer lugar, el problema

\begin{equation}
H_{+} \phi(r,\theta)= \left(
\begin{array}{cc}
 m & ie^{- i\theta}
\left(\partial_{r}-\frac{i}{r}\partial_{\theta}-\frac{eB}{2}r\right)
\\
  ie^{i\theta}
\left(\partial_{r}+\frac{i}{r}\partial_{\theta}+\frac{eB}{2}r\right)
& -m
\end{array}
\right)\phi(r,\theta)=E\phi(r,\theta).
\end{equation}

Dado que vamos a obtener una base com\'{u}n de autofunciones de los operadores
hamiltoniano y componente $z$ del momento angular total, proponemos a $\phi$
autofunci\'{o}n de este \'{u}ltimo

\be
\phi_{k}=\left(\begin{array}{c}
  \ f_{k}(r)e^{ik\theta } \\
  \ g_{k}(r)e^{i(k+1)\theta }
\end{array}\right)\quad ,\textrm{con}\quad k\in\mathbf{Z}.
\ee

Entonces, se satisface el siguiente sistema de ecuaciones de
primer orden acoplado

\be
i\left(\frac{d}{dr}+\frac{k+1}{r}-\frac{eB}{2}r\right)g_{k}(r)
=(E-m) f_{k}(r) \ee
\be
i\left(\frac{d}{dr}-\frac{k}{r}+\frac{eB}{2}r\right)f_{k}(r) =(E+m)g_{k}(r) , \ee o
bien, definiendo la variable adimensional $\rho=\left(\frac{eB}{2}\right)^{1/2}r$

\be
i\left(\frac{eB}{2}\right)^{1/2}
\left(\frac{d}{d\rho}+\frac{k+1}{\rho}-\rho\right)g_{k}(\rho) =(E-m) f_{k}(\rho)
\label{fg1}\ee
\be
i\left(\frac{eB}{2}\right)^{1/2}
\left(\frac{d}{d\rho}-\frac{k}{\rho}+\rho\right)f_{k}(\rho)
=(E+m)g_{k}(\rho)\label{fg2} .\ee

\bigskip

Analizaremos, en primer lugar, el problema $E\neq -m$. En este
caso deberemos resolver

\begin{equation}\label{f}
\left[-\frac{1}{\rho}\frac{d}{d\rho}\left(\rho\frac{d}{d\rho}\right)
+\frac{k^{2}}{\rho^{2}}+\rho^{2}\right]f_{k}(\rho)=
\left(\frac{2}{eB}\right)\left[E^{2}-m^{2}+eB(k+1)\right]f_{k}(\rho),
\end{equation}
\be\label{g} g_{k}(\rho)=\frac{i}{E+m}\left(\frac{eB}{2}\right)^{1/2}
\left(\frac{d}{d\rho}-\frac{k}{\rho}+\rho\right)f_{k}(\rho). \ee

A fin de resolver la ecuaci\'{o}n de segundo orden, hacemos el
siguiente cambio de variables

\begin{equation}
\xi=\rho^{2}\quad \Rightarrow\quad d\xi=2\rho d\rho\quad \Rightarrow\quad
\frac{1}{\rho}\frac{d}{d\rho}=2\frac{d}{d\xi}\quad \textrm{y}\quad
\rho\frac{d}{d\rho}=2\xi\frac{d}{d\xi},
\end{equation}
y proponemos soluciones de la forma

\begin{equation}
 f_{k}\left(\rho(\xi)\right)=e^{-\frac{\xi}{2}}\xi^{\frac{k}{2}}F(\xi).
\end{equation}

Al reemplazarlas en la ecuaci\'{o}n diferencial, obtenemos

\bea \left(\frac{2}{eB}\right)
\left[E^{2}-m^{2}+eB(k+1)\right]e^{-\frac{\xi}{2}}\xi^{\frac{k}{2}}F(\xi)
&=&\left[-4\frac{d}{d\xi}\left(\xi\frac{d}{d\xi}\right)+\frac{k^{2}}{\xi}+\xi\right]
e^{-\frac{\xi}{2}}\xi^{\frac{k}{2}}F(\xi)\nn \\&&\nn\\
&=&\left(-4\xi\frac{d^{2}}{d\xi^{2}}-4\frac{d}{d\xi}+\frac{k^{2}}{\xi}+\xi\right)
e^{-\frac{\xi}{2}}\xi^{\frac{k}{2}}F(\xi) .\eea

Dado que

\begin{equation}
  \frac{d}{d\xi}\left(e^{-\frac{\xi}{2}}\xi^{\frac{k}{2}}F(\xi)\right)=e^{-\frac{\xi}{2}}
  \left(-\frac{\xi^{\frac{k}{2}}F(\xi)}{2}
  +\frac{k\xi^{\frac{k}{2}-1}F(\xi)}{2}\xi^{\frac{k}{2}}\frac{dF}{d\xi}(\xi)\right),
\end{equation}

\begin{equation}
  \frac{d^{2}}{d\xi^{2}}\left(e^{-\frac{\xi}{2}}\xi^{\frac{k}{2}}F(\xi)\right)
  =e^{-\frac{\xi}{2}}\xi^{\frac{k}{2}}
  \left[\frac{d^{2}F}{d\xi^{2}}+\left(\frac{k}{\xi}-1\right)\frac{dF}{d\xi}
  +\left(\frac{1}{4}-\frac{k}{2\xi}+
  \frac{\frac{k}{2}\left(\frac{k}{2}-1\right)}{\xi^{2}}\right)\right].
\end{equation}

\newpage

$F(\xi)$ satisface la ecuaci\'{o}n diferencial

\begin{equation}
  \xi\frac{d^{2}F}{d\xi^{2}}+(k+1-\xi)\frac{dF}{d\xi}+\frac{E^{2}-m^{2}}{2eB}F(\xi)=0,
\end{equation}
que puede identificarse con la ecuaci\'{o}n de Kummer \cite{Abram}.

\begin{equation}
  z\frac{d^{2}w}{dz^{2}}+(b-z)\frac{dw}{dz}-aw=0,
\end{equation}
donde $b=k+1$ y $a=\frac{m^{2}-E^{2}}{2eB}$.

\medskip

Las soluciones de esta ecuaci\'{o}n, estudiadas detalladamente en
el Ap\'{e}ndice A, conducen a autofunciones de cuadrado integrable
s\'{o}lo en los casos en que $a$ es un entero negativo o cero, y
$b\geq a+1$. Estas soluciones son polinomios generalizados de
Laguerre. Cuando $a=-n$ los autovalores permitidos para la
energ\'{\i}a deben satisfacer

\begin{equation}
E^{2}=m^{2}+2eBn\quad \textrm{con}\quad n=0, 1, 2, ...
\end{equation}
mientras que las soluciones $F(\xi)$ resultan

\begin{equation}
  F(\xi)=\left\{\begin{array}{ll}
    C_{k,n}.L_{n}^{k}(\xi)& \textrm{si $k \geq 0$}\\
    C_{k,n}.\xi^{-k}L_{n+k}^{-k}(\xi)&\textrm{si $-n \leq k <0$}\
  \end{array}\right. ,
\end{equation}
donde $C_{k,n}$ es una constante compleja que se ajusta en cada caso para que los
estados resulten de norma 1. Entonces, las soluciones de la ecuaci\'{o}n (\ref{f}) son

\begin{equation}
  f_{k}(\rho)=\left\{\begin{array}{ll}
    C_{k,n}.e^{-\frac{\rho^{2}}{2}}\rho^{k}L_{n}^{k}(\rho^{2})
    & \textrm{si $k \geq 0$}\\
    C_{k,n}.e^{-\frac{\rho^{2}}{2}}\rho^{-k}L_{n+k}^{-k}(\rho^{2})
    &\textrm{si $-n \leq k <0$}\
  \end{array}\right. .
\end{equation}

Usando relaciones de recurrencia entre los polinomios generalizados de Laguerre
\cite{Abram}, es f\'{a}cil obtener las soluciones de la ecuaci\'{o}n (\ref{g}), para
$n\neq 0$

\begin{equation}
  g_{k}(\rho)=\left\{\begin{array}{ll}
    -iC_{k,n}\frac{(2eB)^{\frac{1}{2}}}{E+m}e^{-\frac{\rho^{2}}{2}}\rho^{k+1}L_{n-1}^{k+1}(\rho^{2})
  & \textrm{si $k \geq 0$}\\
    iC_{k,n}\frac{(2eB)^{\frac{1}{2}}}{E+m}n.e^{-\frac{\rho^{2}}{2}}\rho^{-k-1}L_{n+k}^{-k-1}(\rho^{2})
  &\textrm{si $-n \leq k <0$}\
  \end{array}\right. ,
\end{equation}
y $g_{k}(\rho)=0$ para $n=0$.

\newpage

Ahora analizaremos el caso $E=-m$. En este caso, la ecuaci\'{o}n
(\ref{fg2}) queda desacoplada

\be
\left(\frac{d}{d\rho}-\frac{k}{\rho}+\rho\right)f_{k}(\rho)=0, \ee y su soluci\'{o}n es
simplemente

\be
f_{k}(\rho)=C_{k}.e^{-\frac{\rho^{2}}{2}}\rho^{k}, \ee donde $k$ solamente puede tomar
valores enteros positivos para que la componente superior de $\phi$ resulte de cuadrado
integrable, y $C_{k}$ es una constante de normalizaci\'{o}n.

Por su parte, la ecuaci\'{o}n (\ref{fg1}) es una ecuaci\'{o}n de primer orden
inhomog\'{e}nea

\be
\left(\frac{d}{d\rho}+\frac{k+1}{\rho}-\rho\right)g_{k}(\rho)
=\frac{2miC_{k}}{\left(\frac{eB}{2}\right)^{1/2}}e^{-\frac{\rho^{2}}{2}}\rho^{k}.
\label{inhom}\ee

Proponemos soluciones de la forma

\be
g_{k}(\rho)=G(\rho)e^{\frac{\rho^{2}}{2}}\rho^{-(k+1)}, \ee por lo que $G(\rho)$ debe
satisfacer

\be
\frac{dG}{d\rho}=\frac{2miC_{k}}{\left(\frac{eB}{2}\right)^{1/2}}
e^{-\rho^{2}}\rho^{2k+1},\ee

\be
G(\rho)=\frac{2miC_{k}}{\left(\frac{eB}{2}\right)^{1/2}}\int d\rho
e^{-\rho^{2}}\rho^{2k+1}, \ee y haciendo el cambio de variables $\rho^{2}=x$

\bea G(\rho(x))&=&\frac{miC_{k}}{\left(\frac{eB}{2}\right)^{1/2}}\int dx
e^{-x}x^{k}\nn\\
&=&-\frac{miC_{k}}{\left(\frac{eB}{2}\right)^{1/2}}e^{-x}\sum_{n=0}^{k}\frac{k!}{(k-n)!}x^{k-n}\,+Cte.
\eea

Por lo tanto, la soluci\'{o}n general de (\ref{inhom}) es la soluci\'{o}n general de su
homog\'{e}nea corres\-pondiente m\'{a}s una soluci\'{o}n particular de la
inhomog\'{e}nea

\be
g_{k}(\rho)=-\frac{miC_{k}}{\left(\frac{eB}{2}\right)^{1/2}}e^{-\frac{\rho^{2}}{2}}
\sum_{n=0}^{k}\frac{k!}{(k-n)!}\rho^{k-2n-1} + De^{\frac{\rho^{2}}{2}}\rho^{-(k+1)},\ee
la cual conduce, para todo valor de de la constante $D$, a autofunciones de la
componente inferior de $\phi$ que no son de cuadrado integrable.

\bigskip

Por lo tanto, los autovalores de $H_{+}$ que corresponden a autofunciones de cuadrado
integrable son los siguientes:
\be
E_{+}=\left\{\begin{array}{l}
  m \\
  \pm \sqrt{m^{2}+2eBn}\quad\textrm{con}\quad n=1,2,3,...
\end{array}\right.\label{E+}\ee

\bigskip

El problema

\begin{equation}
H_{-} \chi(r,\theta)= \left(
\begin{array}{cc}
 m & ie^{ i\theta}
\left(\partial_{r}+\frac{i}{r}\partial_{\theta}+\frac{eB}{2}r\right)
\\
  ie^{-i\theta}
\left(\partial_{r}-\frac{i}{r}\partial_{\theta}-\frac{eB}{2}r\right)
& -m
\end{array}
\right)\chi(r,\theta)=E\chi(r,\theta),
\end{equation}
se resuelve en forma enteramente an\'{a}loga.

En efecto, factorizando la dependencia angular, obtenemos un sistema de ecuaciones de
primer orden acopladas similar al de (\ref{fg1}) y (\ref{fg2}), que puede desacoplarse
en una ecuaci\'{o}n de segundo orden para el caso $E\neq m$. Nuevamente, los
autovalores de la energ\'{\i}a satisfacen

\begin{equation}
E^{2}=m^{2}+2eBn\quad \textrm{con}\quad n=0, 1, 2, ...
\end{equation}
y las autofunciones pueden expresarse en t\'{e}rminos de polinomios generalizados de
Laguerre.

\medskip

Por \'{u}ltimo, al estudiar las soluciones del caso $E=m$, se
encuentra que este autovalor no tiene asociadas autofunciones de
cuadrado integrable.

Finalmente, nos quedan los siguientes autovalores de $H_{-}$ correspondientes a
autofunciones de cuadrado integrable:

\be
E_{-}=\left\{\begin{array}{l}
  -m \\
  \pm \sqrt{m^{2}+2eBn}\quad\textrm{con} \quad n=1,2,3,...
\end{array}\right.\label{E-}\ee

\newpage

\section{Energ\'{i}a de vac\'{i}o en $2+1$ dimensiones}
\label{vacio-2+1}

Como se explic\'{o} en la secci\'{o}n I, la energ\'{\i}a de vac\'{\i}o est\'{a} dada
por la ecuaci\'{o}n (\ref{E0})

\begin{equation}
E_{0}=-\frac{1}{2}\sum_{n}g_{n}.|E_{n}| ,
\end{equation}
donde los $E_{n}$ son los autovalores dados por las ecuaciones (\ref{E+}) y (\ref{E-}).
El factor de dege\-neraci\'{o}n $g_{n}$ proviene de los distintos valores que puede
tomar $k$ (cualquier entero mayor o igual que n). \'{E}ste puede calcularse contando el
n\'{u}mero de estados que se encuentran en el problema de la part\'{\i}cula libre con
m\'{o}dulo cuadrado del impulso lineal entre $2eBn$ y $2eB(n+1)$, si pensamos que al
encender el campo todos estos niveles colapsan en uno. Integrando la densidad de
estados de la part\'{\i}cula libre, resulta

\begin{equation}
g_{n}=\frac{L_{x}L_{y}}{(2\pi)^{2}}\int\int
dp_{x}dp_{y}=\frac{A}{(2\pi)^{2}}2\pi\int_{\sqrt{2eBn}}^{\sqrt{2eB(n+1)}}pdp
=\frac{A}{2\pi}eB ,
\end{equation}
donde $A$ es el \'{a}rea perpendicular al campo magn\'{e}tico.

\medskip

Conviene, entonces, considerar la energ\'{\i}a de vac\'{\i}o por unidad de \'{a}rea

\begin{equation}
\varepsilon_{0}=\frac{E_{0}}{A}=-\frac{eB}{4\pi}\sum_{n}|E_{n}|=
-\frac{eB}{4\pi}2\left(m+2\sum_{n=1}^{\infty}\sqrt{m^{2}+2eBn}\right).
\end{equation}

Como es usual, este resultado es s\'{o}lo formal y para darle sentido
debemos regularizarlo. Hacemos esto con el m\'{e}todo de
regularizaci\'{o}n de la funci\'{o}n zeta, descrito en la secci\'{o}n I.
Debemos multiplicar a la suma por un par\'{a}metro para que conserve
unidades de energ\'{\i}a por unidad de \'{a}rea. Usamos $\mu$ con unidades
de masa

\bea \varepsilon_{0}&=&-\frac{2eB}{4\pi}\left\{\mu^{s+1}
\left[m^{-s}+2\sum_{n=1}^{\infty}\left(m^{2}+2eBn\right)^{-s/2}\right]\right\}_{s=-1}\nn
\\&&\nn\\ &=&\frac{2eB}{2\pi}\left\{\mu^{s+1}\left[\frac{m^{-s}}{2}-(2eB)^{-s/2}
\sum_{n=0}^{\infty}\left(n+\frac{m^{2}}{2eB}\right)^{-s/2}\right]\right\}_{s=-1}.\eea

Es conveniente introducir los par\'{a}metros adimensionales

\be\label{paradi} \left\{\begin{array}{l}
  b=\frac{2eB}{\mu^{2}} \\
  c=\frac{m^{2}}{\mu^{2}}
\end{array}\right. ,\ee
de modo que

\begin{equation}\label{E0bc}
\varepsilon_{0}=\frac{b}{2\pi}\mu^{3} \left[\frac{c^{-s/2}}{2}
-b^{-s/2}\sum_{n=0}^{\infty}\left(n+\frac{c}{b}\right)^{-s/2}\right]_{s=-1},
\end{equation}

Para $Re(s)>2$, la serie en la ecuaci\'{o}n (\ref{E0bc}) converge y coincide con la
funci\'{o}n zeta Hurwitz.

\begin{equation}\label{Hurwitz}
  \zeta_{H}(s,a)=\sum_{n=0}^{\infty}(n+a)^{-s},
\end{equation}
que es convergente para $Re(s)>1$, pero se extiende anal\'{\i}ticamente a una
funci\'{o}n meromorfa con un polo simple en $s=1$. Por lo tanto, la densidad de
energ\'{\i}a de vac\'{\i}o puede definirse, por extensi\'{o}n anal\'{\i}tica, como

\begin{equation}
\varepsilon_{0}=\frac{b}{2\pi}\mu^{3}
\left[\frac{c^{-s/2}}{2}-(b)^{-s/2}\zeta_{H}\left(s/2,\frac{c}{b}\right)\right]_{s=-1},
\end{equation}

Dado que $\zeta_{H}\left(s/2,\frac{c}{b}\right)$ tiene un \'{u}nico polo simple en
$s=2$, la densidad de energ\'{\i}a de vac\'{\i}o por unidad de \'{a}rea resulta finita
e igual a

\begin{equation}\label{vac2}
\varepsilon_{0}=\frac{2eB}{2\pi}
\left[\frac{m}{2}-(2eB)^{1/2}\zeta_{H}\left(-1/2,\frac{m^{2}}{2eB}\right)\right].
\end{equation}

Vale la pena se\~{n}alar que, en el caso de un campo puramente magn\'{e}tico, la
energ\'{\i}a de vac\'{\i}o debe coincidir con la densidad lagrangiana efectiva al orden
1 loop del sistema. En efecto, el resultado de la ecuaci\'{o}n (\ref{vac2}) coincide
con la densidad lagrangiana efectiva obtenida en la referencia \cite{Blau}, a partir
del c\'{a}lculo con regularizaci\'{o}n zeta del determinante del operador de Dirac.

Para calcular la densidad de energ\'{\i}a de Casimir, debemos conocer la densidad de
energ\'{\i}a de vac\'{\i}o del problema sin campo. A fin de evaluar (\ref{vac2}) para
$B=0$, usaremos un desarrollo asint\'{o}tico de la funci\'{o}n $\zeta_{H}(s,a)$,
v\'{a}lido cuando $|a|\rightarrow\infty$ \cite{Bateman}.

\bea\label{zetades}
\zeta_{H}(s,a)&=&\frac{1}{\Gamma(s)}\left[a^{-(s-1)}\Gamma(s-1)+\frac{1}{2}a^{-s}\Gamma(s)\right.
+{}\nn\\&&\nn\\
&&\left.{}+\sum_{n=1}^{m}\frac{B_{2n}}{(2n)!}a^{-(2n+s-1)}\Gamma(2n+s-1)+O(a^{-(2m+s+1)})\right],
\eea donde $B_{2n}$ son los coeficientes de Bernoulli.
\bigskip

Al escribir la densidad de energ\'{\i}a de vac\'{\i}o en t\'{e}rminos de este
desarrollo asint\'{o}tico resulta, para $eB\ll m^{2}$

\begin{equation}
  \varepsilon_{0}=\frac{m^{3}}{3\pi}-\frac{m^{3}}{2\pi}\frac{1}{\Gamma(-1/2)}
\sum_{n=1}^{m}\frac{B_{2n}}{(2n)!}\left(\frac{2eB}{m^{2}}\right)^{2n}\Gamma(2n-3/2)
+O\left[\left(\frac{2eB}{m^{2}}\right)^{2(m+1)}\right] \label{vac2des}.\end{equation}

Vemos que el primer t\'{e}rmino, independiente del campo magn\'{e}tico, es
$\frac{m^{3}}{3\pi}$, y por lo tanto corresponde a la densidad de
energ\'{\i}a de vac\'{\i}o del problema sin campo. En el Ap\'{e}ndice B
obtenemos el mismo resultado sumando directamente los autovalores
del espectro cl\'{a}sico de energ\'{\i}as de la part\'{\i}cula libre.

\bigskip

Resumiendo, la densidad de energ\'{\i}a de Casimir resulta, para valores arbitrarios
del campo

\bea \varepsilon_{Cas}&=&\varepsilon_{0}-\varepsilon_{0}^{l}\nn\\&&\nn\\
&=&\frac{2eB}{2\pi}
\left[\frac{m}{2}-(2eB)^{1/2}\zeta_{H}\left(-1/2,\frac{m^{2}}{2eB}\right)\right]
-\frac{m^{3}}{3\pi} .\eea

\bigskip

N\'{o}tese que, para campos peque\~{n}os, el segundo t\'{e}rmino del desarrollo
(\ref{zetades}) cancela, en la ecuaci\'{o}n (\ref{vac2}), el t\'{e}rmino proporcional
al campo. De este modo, resulta para la energ\'{\i}a de Casimir un desarrollo en
potencias pares del campo magn\'{e}tico.

\newpage

\section{Espectro en $3+1$ dimensiones}
\label{esp-3+1}

En lugar de resolver expl\'{\i}citamente (\ref{H4}) podemos aplicar una
transformaci\'{o}n de Lorentz sobre las soluciones independientes de $z$. En
particular, considaremos un boost en la direcci\'{o}n del eje $z$, desde el sistema de
referencia en el que $p_{z}=0$ hasta el sistema en el cual $p_{z}'\neq 0$

\be{\Lambda^{\mu}}\nu=\left(\begin{array}{cccc}
  \cosh \omega & 0 & 0 & \sinh \omega\\
  0 & 1 & 0 & 0 \\
  0 & 0 & 1 & 0 \\
  \sinh \omega & 0 & 0 & \cosh \omega
\end{array} \right),\qquad \textrm{donde}\quad \cosh \omega=\frac{1}{\sqrt{1-v^{2}}}.\ee

Entonces

\be\left\{\begin{array}{cc}
  E' & =\cosh\omega E \\
  p_{z}' & =\sinh\omega E
\end{array}\right. , \ee
por lo tanto

\be {E'}^{2}=\cosh^{2}\omega E^{2}=(1+\sinh^{2}\omega)E^{2}=E^{2}+ {p_{z}'}^{2}.\ee

Por su parte, los espinores se transforman seg\'{u}n

\be
\Psi'=S\Psi\quad ,\ee donde

\be
S=\exp\left(-i\frac{\omega}{2}\,\frac{i}{2}[\gamma^{0},\gamma^{3}]\right)
=\cosh\left(\frac{\omega}{2}\right) {\mathbf{1}}_{4}
+\sinh\left(\frac{\omega}{2}\right)\gamma^{0}\gamma^{3}.\ee

En la representaci\'{o}n elegida para las matrices de Dirac resulta

\be
S=\left(\begin{array}{cc}
  \cosh\left(\frac{\omega}{2}\right)\mathbf{1}_{2}
  & -\sinh\left(\frac{\omega}{2}\right)\sigma_{2} \\
  -\sinh\left(\frac{\omega}{2}\right)\sigma_{2} &
  \cosh\left(\frac{\omega}{2}\right)\mathbf{1}_{2}
\end{array}\right).\ee

A partir de las soluciones

\be
\Psi=\left(\begin{array}{c}
  \phi(r,\theta) \\
  0
\end{array}\right)e^{-iE_{+}t},\ee
donde $\phi$ es una de las autofunciones del bloque $H_{+}$, se obtiene

\be
\Psi'=\left(\begin{array}{c}
  \cosh\left(\frac{\omega}{2}\right)\phi(r,\theta) \\
  -\sinh\left(\frac{\omega}{2}\right)\sigma_{2}\phi(r,\theta)
\end{array}\right)e^{-iE_{+}t} , \ee
o bien, en t\'{e}rminos de las coordenados del sistema primado,

\be
\Psi'=\left(\begin{array}{c}
  \cosh\left(\frac{\omega}{2}\right)\phi(r',\theta') \\
  -\sinh\left(\frac{\omega}{2}\right)\sigma_{2}\phi(r',\theta')
\end{array}\right)e^{ip_{z}'z'}e^{-iE_{+}'t'}.\ee

Estas autofunciones corresponden a los autovalores de la energ\'{\i}a

\be
E_{+}'=\left\{\begin{array}{l}
  \sqrt{m^{2} +{p_{z}'}^{2}}\\
  \pm \sqrt{m^{2}+{p_{z}'}^{2}+2eBn}
\end{array}\right.\quad\textrm{con}\quad p_{z}'\in {\mathbf{R}}
\quad \textrm{y}\quad n=1,2,3,...\label{E+3}\ee

Similarmente, a partir de las autofunciones $\chi$ del bloque
$H_{-}$, se obtienen los siguientes espinores en el sistema de
referencia primado

\be
\Psi'=\left(\begin{array}{c}
  -\sinh\left(\frac{\omega}{2}\right)\sigma_{2}\chi(r',\theta') \\
\cosh\left(\frac{\omega}{2}\right)\chi(r',\theta')
\end{array}\right)e^{ip_{z}'z'}e^{-iE_{-}'t'} , \ee
que corresponden a autovalores

\be
E_{-}'=\left\{\begin{array}{l}
 - \sqrt{m^{2} +{p_{z}'}^{2}}\\
  \pm \sqrt{m^{2}+{p_{z}'}^{2}+2eBn}
\end{array}\right.\quad\textrm{con}\quad p_{z}'\in {\mathbf{R}}
\quad \textrm{y}\quad n=1,2,3,...\label{E-3}\ee

\newpage

\section{Energ\'{i}a de vac\'{i}o en $3+1$ dimensiones}
\label{vacio-3+1}

En este caso, la energ\'{\i}a de vac\'{\i}o se obtiene a partir de sumar los
autovalores dados en la ecuaciones (\ref{E+3}) y (\ref{E-3})

\begin{equation}
  E_{0}=-g\frac{L_{z}}{2\pi}\int_{-\infty}^{\infty}dp_{z}
  \left(\sqrt{m^{2}+p_{z}^{2}}+
  2\sum_{n=1}^{\infty}\sqrt{m^{2}+p_{z}^{2}+2eBn}\right),
\end{equation}
donde el factor de degeneraci\'{o}n $g$ es nuevamente $\frac{A}{2\pi}eB$.

\medskip

Redefinimos el \'{\i}ndice de la suma para que \'{e}sta comience desde 0 y definimos la
densidad de energ\'{\i}a de vac\'{\i}o por unidad de volumen

\begin{equation}
\varepsilon_{0}=\frac{E_{0}}{AL_{z}}=\frac{eB}{(2\pi)^{2}}\int_{-\infty}^{\infty}dp_{z}
  \left(\sqrt{m^{2}+p_{z}^{2}}-
  2\sum_{n=0}^{\infty}\sqrt{m^{2}+p_{z}^{2}+2eBn}\right).
\end{equation}

Nuevamente, la expresi\'{o}n que encontramos para la densidad de energ\'{\i}a de
vac\'{\i}o es formalmente divergente. Repitiendo el procedimiento antes utilizado, la
extendemos anal\'{\i}ticamente con la t\'{e}cnica de regularizaci\'{o}n zeta. Como
antes debemos agregar alg\'{u}n par\'{a}metro para que la densidad de energ\'{\i}a de
vac\'{\i}o mantenga las unidades apropiadas. Usando $\mu$ con unidades de masa, resulta

\be\label{E0regu}
  \varepsilon_{0}=\frac{eB}{(2\pi)^{2}}\left\{\mu^{s+1}\int_{-\infty}^{\infty}dp_{z}
  \left[(m^{2}+p_{z}^{2})^{-s/2}-2\sum_{n=0}^{\infty}\left(m^{2}+p_{z}^{2}+2eBn\right)^{-s/2}
  \right]\right\}_{s=-1}.
\end{equation}

Si usamos la transformada de Mellin \cite{blauviss}

\begin{equation}\label{mellin}
 Z^{-s}=\frac{1}{\Gamma(s)}\int_{0}^{\infty}dt t^{s-1}e^{-Zt},
\end{equation}
la ecuaci\'{o}n (\ref{E0regu}) se escribe

\bea \varepsilon_{0}&=&
\frac{eB}{(2\pi)^{2}}\left\{\mu^{s+1}\int_{-\infty}^{\infty}dp_{z}\left[
\frac{1}{\Gamma(s/2)}\int_{0}^{\infty}dtt^{s/2-1}
\exp[-(m^{2}+p_{z}^{2})t]\right.\right. -{}\nn \\&&\nn
\\&&\left.\left.{}-2\sum_{n=0}^{\infty}\frac{1}{\Gamma(s/2)}\int_{0}^{\infty}dt
t^{s/2-1}\exp[-(m^{2}+p_{z}^{2}+2eBn)t]\right]\right\}_{s=-1} .\eea

El orden de integraci\'{o}n puede intercambiarse para $Re(s)$ suficientemente grande,
de modo que

\bea \varepsilon_{0}&=&\frac{eB}{(2\pi)^{2}}\left\{\mu^{s+1}
\left[\frac{1}{\Gamma(s/2)}\int_{0}^{\infty}dtt^{s/2-1}\exp(-m^{2}t)\int_{-\infty}^{\infty}dp_{z}
\exp(-p_{z}^{2}t)\right.\right. -{}\nn\\ &&\nn\\
&&\left.\left.{}-2\sum_{n=0}^{\infty}\frac{1}{\Gamma(s/2)}\int_{0}^{\infty}dt
t^{s/2-1}\exp[-(m^{2}+2eBn)t]\int_{-\infty}^{\infty}dp_{z}\exp(-p_{z}^{2}t)
\right]\right\}_{s=-1} .\eea

Usando el resultado de la siguiente integral definida

\begin{equation}\label{int}
\int_{-\infty}^{\infty}dx e^{-ax^{2}}=\sqrt{\frac{\pi}{a}},
\end{equation}
resolvemos las integrales a lo largo de $p_{z}$. As\'{\i} obtenemos

\bea \label{E0cita}\varepsilon_{0} &=&\frac{eB}{(2\pi)^{2}}\left\{\mu^{s+1}
\left[\frac{\sqrt{\pi}}{\Gamma(s/2)}\int_{0}^{\infty}dtt^{\frac{s-1}{2}-1}\exp(-m^{2}t)
\right.\right.-{}\nn\\ &&\nn\\
&&\left.\left.{}-2\sum_{n=0}^{\infty}\frac{\sqrt{\pi}}{\Gamma(s/2)}\int_{0}^{\infty}dt
t^{\frac{s-1}{2}-1}\exp\left[-(m^{2}+2eBn)t\right]\right]\right\}_{s=-1}. \eea

Luego de realizar las integrales en $t$, resulta

\be \varepsilon_{0} =\frac{eB}{(2\pi)^{2}}\left\{\mu^{s+1}\left[
\sqrt{\pi}\frac{\Gamma(\frac{s-1}{2})}{\Gamma(s/2)}(m^{2})^{-\frac{s-1}{2}}
-2\sqrt{\pi}\frac{\Gamma(\frac{s-1}{2})}{\Gamma(s/2)}
\sum_{n=0}^{\infty}(m^{2}+2eBn)^{-\frac{s-1}{2}} \right]\right\}_{s=-1}, \ee

Nuevamente, la serie en el segundo t\'{e}rmino define, para $Re(s)>3$, una zeta de
Hurwitz. Por extensi\'{o}n anal\'{\i}tica, obtenemos

\be\label{E0ext} \varepsilon_{0}=
\frac{2eB}{(2\pi)^{2}}\sqrt{\pi}\left\{\mu^{s+1}\frac{\Gamma(\frac{s-1}{2})}{\Gamma(s/2)}
\left[\frac{(m^{2})^{-\frac{s-1}{2}}}{2}
-(2eB)^{-\frac{s-1}{2}}\zeta_{H}\left(\frac{s-1}{2},\frac{m^{2}}{2eB}\right)
\right]\right\}_{s=-1}.\ee

A diferencia de lo que ocurr\'{\i}a en $2+1$ dimensiones, la extensi\'{o}n de la
funci\'{o}n zeta del hamiltoniano cl\'{a}sico presenta, en $3+1$, un polo simple en el
valor $s=-1$, debido al comportamiento de la funci\'{o}n $\Gamma(s)$ en los enteros
negativos.

Usando nuevamente los par\'{a}metros adimensionales definidos en (\ref{paradi}),
(\ref{E0ext}) puede escribirse

\bea \label{E0adi}\varepsilon_{0}&=&\frac{\sqrt{\pi}}{(2\pi)^{2}}b\mu^{4}
\left\{\frac{\Gamma(\frac{s-1}{2})}{\Gamma(s/2)}
\left[\frac{c^{-\frac{s-1}{2}}}{2}-b^{-\frac{s-1}{2}}
\zeta_{H}\left(\frac{s-1}{2},\frac{c}{b}\right)\right]\right\}_{s=-1}\nn\\ &&\nn\\
&=&\frac{\sqrt{\pi}}{\pi^{2}}b\mu^{4}\left\{\frac{\Gamma(\frac{s+3}{2})}{\Gamma(s/2)(s+1)(s-1)}
\left[\frac{c^{-\frac{s-1}{2}}}{2}-b^{-\frac{s-1}{2}}
\zeta_{H}\left(\frac{s-1}{2},\frac{c}{b}\right)\right]\right\}_{s=-1} ,\eea o bien

\begin{equation}\left.
\varepsilon_{0}=\frac{\sqrt{\pi}}{\pi^{2}}b\mu^{4}\frac{Z(s)}{s+1}\right|_{s=-1},
\end{equation}
donde

\begin{equation}
  Z(s)=\frac{\Gamma(\frac{s+3}{2})}{\Gamma(s/2)(s-1)}\left[\frac{c^{-\frac{s-1}{2}}}{2}
  -b^{-\frac{s-1}{2}}\zeta_{H}\left(\frac{s-1}{2},\frac{c}{b}\right)\right].
\end{equation}

A fin de identificar las partes singular y finita de la densidad de energ\'{\i}a de
vac\'{\i}o, hacemos un desarrollo alrededor de $s=-1$.

\begin{equation}\label{E0Z}
\varepsilon_{0}=\frac{\sqrt{\pi}}{\pi^{2}}b\mu^{4}\left[\frac{Z(-1)}{s+1}+Z'(-1)\right]_{s=-1}.
\end{equation}

Para la parte singular debemos calcular

\begin{equation}
Z(-1)=\frac{\Gamma(1)}{(-2)\Gamma(-1/2)}\left[\frac{c}{2}-b\
\zeta_{H}\left(-1,\frac{c}{b}\right)\right],
\end{equation}
y usando la extensi\'{o}n anal\'{\i}tica de la funci\'{o}n Zeta de Hurwitz

\begin{equation}\
\zeta_{H}(-1,a)=-\frac{1}{2}\left(a^{2}-a+\frac{1}{6}\right),
\end{equation}

\begin{equation}\label{Z}
Z(-1)=\frac{b}{8\sqrt{\pi}}\left(\frac{c^{2}}{b^{2}}+\frac{1}{6}\right).
\end{equation}

La parte finita est\'{a} dada por

\bea\label{Zprima} Z'(-1)&=&\frac{b}{16\sqrt{\pi}}\left(\psi(1)-\psi(-1/2)+1\right)
\left[\frac{c}{b}-2\zeta_{H}\left(-1,\frac{c}{b}\right)\right]-{}\nn\\ &&\nn\\
&&{}-\frac{b}{16\sqrt{\pi}}\left(\frac{c}{b}\ln c -2\ln b\
\zeta_{H}\left(-1,\frac{c}{b}\right)+2\zeta_{H}'\left(-1,\frac{c}{b}\right)\right)\nn\\
&&\nn\\ &=&\frac{b}{16\sqrt{\pi}}\left(\psi(1)-\psi(-1/2)+1\right)
\left(\frac{c^{2}}{b^{2}}+\frac{1}{6}\right)-{}\nn\\ &&\nn\\
&&{}-\frac{b}{16\sqrt{\pi}}\left[\frac{c}{b}\ln c+\ln b\left(\frac{c^{2}}{b^{2}}
-\frac{c}{b}+\frac{1}{6}\right)+2\zeta_{H}'\left(-1,\frac{c}{b}\right)\right]. \eea

Reemplazando (\ref{Z}) y (\ref{Zprima}) en (\ref{E0Z}), la densidad de energ\'{\i}a de
vac\'{\i}o resulta

\bea\label{E0polo}
\varepsilon_{0}&=&\mu^{4}\left[\frac{c^{2}}{8\pi^{2}(s+1)}+\frac{\frac{b^{2}}{6}}{8\pi^{2}(s+1)}
\right]_{s=-1}+{}\nn\\ &&\nn\\
&&{}+\frac{\mu^{4}}{16\pi^{2}}\left[(\psi(1)-\psi(-1/2)+1-\ln c)
\left(c^{2}+\frac{b^{2}}{6}\right)\right.+{} \nn\\ &&\nn\\ &&\left.{} +\left(c^{2}-c\ b
+\frac{b^{2}}{6}\right)\ln\left(\frac{c}{b}\right)
-2b^{2}\zeta_{H}'\left(-1,\frac{c}{b}\right)\right] .\eea

Vale la pena se\~{n}alar, en este punto, que la parte finita de la expresi\'{o}n
anterior coincide, a menos de t\'{e}rminos proporcionales a $b^{2}$ (y, como veremos,
renormalizables) con la densidad lagrangiana efectiva calculada en \cite{Blau} a partir
del determinante del operador de Dirac. La diferencia entre ambos resultados est\'{a}
de acuerdo con la discusi\'{o}n general presentada en \cite{blauviss}.

De la expresi\'{o}n (\ref{E0polo}), vemos que el residuo en el polo presenta un
t\'{e}rmino independiente del campo magn\'{e}tico, y otro proporcional a su cuadrado.

A fin de dar una interpretaci\'{o}n f\'{\i}sica a este resultado verificaremos, en
primer lugar, que el t\'{e}rmino independiente de $b$ se cancela con uno id\'{e}ntico
que aparece en la densidad de energ\'{\i}a de vac\'{\i}o a $B=0$.

Para hallar la densidad de energ\'{\i}a de vac\'{\i}o sin campo $\varepsilon_{0}^{l}$,
podemos considerar un desa\-rrollo de asint\'{o}tico de la derivada de la funci\'{o}n
zeta en el l\'{\i}mite de los campos peque\~{n}os, y quedarnos con los t\'{e}rminos del
resultado final independientes del campo (en el Ap\'{e}ndice C la reobtendremos
directamente a partir de la suma de los autovalores del espectro cl\'{a}sico de la
part\'{\i}cula libre).

El siguiente desarrollo para la derivada de la funci\'{o}n zeta de Hurwitz es
v\'{a}lido en el l\'{\i}mite $a\rightarrow\infty$ \cite{Blau}.

\begin{equation}
\zeta_{H}'(-1,a)=\frac{1}{2}\left(a^{2}-a+\frac{1}{6}\right)\ln a
-\frac{1}{4}a^{2}+\frac{1}{12}+\sum_{n=1}^{m-1}\frac{B_{2n+2}a^{-2n}}{(2n)(2n+1)(2n+2)}
+O(a^{-2m}).
\end{equation}

Haciendo uso del mismo resulta, para $b\ll c$

\bigskip

\bea Z'(-1)&=&\frac{b}{16\sqrt{\pi}}\left\{\left(\psi(1)-\psi(-1/2)+1-\ln c\right)
\left(\frac{c^{2}}{b^{2}}+\frac{1}{6}\right)
+\frac{1}{2}\frac{c^{2}}{b^{2}}-\frac{1}{6}\right.-{}\nn\\&&\nn\\
&&\left.{}-2\sum_{n=1}^{m-1}\frac{B_{2n+2}\left(\frac{c}{b}\right)^{-2n}}{(2n)(2n+1)(2n+2)}
+O\left[\left(\frac{b}{c}\right)^{2m}\right]\right\} \eea

y, por lo tanto,

\be\label{E0libre}
\varepsilon_{0}^{l}=\mu^{4}\left[\frac{c^{2}}{8\pi^{2}(s+1)}\right]_{s=-1}
+\frac{\mu^{4}\ c^{2}}{16\pi^{2}}\left[\psi(1)-\psi(-1/2)+3/2-\ln c\right] .\ee

De este modo, la densidad de energ\'{\i}a de Casimir resulta

\bea
  \varepsilon_{Cas}&=&\varepsilon_{0}-\varepsilon_{0}^{l}\nn\\
&&\nn\\ &=& \mu^{4}\left[\frac{\frac{b^{2}}{6}}{8\pi^{2}(s+1)}\right]_{s=-1}
+\frac{\mu^{4}}{16\pi^{2}}\left[(\psi(1)-\psi(-1/2)+1-\ln c)\frac{b^{2}}{6}
-\frac{c^{2}}{2}\right.+{}\nn\\ &&\nn\\ &&\left.{}+\left(c^{2}-c\ b
+\frac{b^{2}}{6}\right)\ln\left(\frac{c}{b}\right)
-2b^{2}\zeta_{H}'\left(-1,\frac{c}{b}\right)\right] . \eea

Esta expresi\'{o}n presenta a\'{u}n una divergencia, cuya dependencia funcional con el
campo es igual a la que aparece en la densidad cl\'{a}sica de energ\'{\i}a

\be \varepsilon_{cl\acute{a}}=\frac{1}{2}B^{2}=\frac{\mu^{4}b^{2}}{8e^{2}}\ee

A fin de eliminar la singularidad que persiste en la densidad de energ\'{\i}a de
Casimir,  renormalizamos la carga $e$. Para ello, consideramos a los t\'{e}rminos
proporcionales al cuadrado del campo como la correcci\'{o}n cu\'{a}ntica de la densidad
de energ\'{\i}a electromagn\'{e}tica cl\'{a}sica.

De esta manera, definimos la carga renornalizada $e_{R}$ a trav\'{e}s de

\begin{equation}\label{eR}
  \left.\frac{\partial(\varepsilon_{cl\acute{a}}+\varepsilon_{0})}{\partial b^{2}}
  \right|_{b^{2}=0}=\frac{\mu^{4}}{8 e_{R}^{2}}
\end{equation}

Evaluamos esta derivada a partir de la ecuaci\'{o}n (\ref{E0adi}) y obtenemos

\bea
\frac{\partial(\varepsilon_{cl\acute{a}}+\varepsilon_{0})}{\partial
b^{2}}&=&\frac{\mu^{4}}{8 e^{2}}+
\frac{\sqrt{\pi}}{(2\pi)^{2}}\mu^{4}\left\{\frac{\Gamma(\frac{s-1}{2})}{\Gamma(s/2)}
\left[\left(\frac{s-3}{4}\right)b^{-\frac{s+1}{2}}
\zeta_{H}\left(\frac{s-1}{2},\frac{c}{b}\right)\right.\right.-{}\nn\\
&&\nn\\ &&\left.\left.{}- \left(\frac{s-1}{4}\right)c\
b^{-\frac{s+3}{2}} \zeta_{H}\left(\frac{s+1}{2},\frac{c}{b}\right)
+\frac{c^{-\frac{s-1}{2}}b^{-1}}{4}\right]\right\}_{s=-1}. \eea
donde hemos usado la siguiente propiedad de la funci\'{o}n zeta de
Hurwitz

\be
\frac{d\zeta_{H}(s,a)}{da}=-s \zeta_{H}(s+1,a)\ee

Podemos evaluar las zetas de Hurwitz en el l\'{\i}mite $b^{2}\rightarrow 0$ con el
desarrollo (\ref{zetades}). Finalmente, resulta

\begin{equation}
  \left.\frac{\partial(\varepsilon_{cl\acute{a}}+\varepsilon_{0})}{\partial b^{2}}
  \right|_{b^{2}=0}=\left.\frac{\mu^{4}}{8 e^{2}}
  -\frac{\sqrt{\pi}}{8\pi^{2}}\mu^{4}B_{2}\ c^{-\frac{s+1}{2}}
  \frac{\Gamma(\frac{s+1}{2})}{\Gamma(s/2)}\right|_{s=-1}.
\end{equation}

Dado que esta expresi\'{o}n es singular en el l\'{\i}mite $s\rightarrow -1$, igual que
como hicimos antes, separamos las partes finita y singular mediante un desarrollo
alrededor de $s=-1$, con lo cual obtenemos

\bea \left.\frac{\partial(\varepsilon_{cl\acute{a}}+\varepsilon_{0})}{\partial b^{2}}
  \right|_{b^{2}=0}&=&\frac{\mu^{4}}{8 e_{R}^{2}}\nn\\ &&\nn\\
  &=&\frac{\mu^{4}}{8 e^{2}}+\frac{\mu^{4}}{96\pi^{2}}
  \left[\frac{2}{s+1}+\psi(1)-\psi(-1/2)-\ln c\right]_{s=-1}, \eea

y de aqu\'{\i} se tiene

\bea
e_{R}^{2}&=&\frac{e^{2}}{1+\frac{e^{2}}{12\pi^{2}}\left[\frac{2}{s+1}+\psi(1)-\psi(-1/2)-\ln
c\right]}\nn\\ &&\nn\\ &\simeq&
e^{2}\left\{1-\frac{e^{2}}{12\pi^{2}}\left[\frac{2}{s+1}+\psi(1)-\psi(-1/2)-\ln
c\right]\right\} .\eea

Entonces, el par\'{a}metro que aparece en el hamiltoniano debe ser renormalizado
seg\'{u}n:

\be
e^{2}=e_{R}^{2}+\frac{e_{R}^{4}}{12\pi^{2}}\left[\frac{2}{s+1}+\psi(1)-\psi(-1/2)-\ln
c\right]+ O(e_{R}^{5}). \ee

Finalmente, para la densidad de energ\'{\i}a al orden 1 loop:

\bea
  \varepsilon_{Cas}&=&\varepsilon_{cl\acute{a}}+\varepsilon_{0}-\varepsilon_{0}^{l}\nn\\
&&\nn\\ &=& \frac{\mu^{4}b^{2}}{8e_{R}^{2}}
+\frac{\mu^{4}}{16\pi^{2}}\left[\frac{b^{2}}{6}-\frac{c^{2}}{2}+\left(c^{2}-c\ b
+\frac{b^{2}}{6}\right)\ln\left(\frac{c}{b}\right)
-2b^{2}\zeta_{H}'\left(-1,\frac{c}{b}\right)\right] , \eea y en t\'{e}rminos de los
par\'{a}metros originales

\bea
  \varepsilon_{Cas}&=& \frac{B^{2}}{2}
+\frac{1}{16\pi^{2}}\left\{\frac{(2e_{R}B)^{2}}{6}\left[\ln
\left(\frac{m^{2}}{2e_{R}B}\right) +1\right]
+m^{4}\left[\ln\left(\frac{m^{2}}{2e_{R}B}\right)-\frac{1}{2}\right]\right.-{}\nn\\
&&\nn\\ &&\left.{}-m^{2}2e_{R}B \ln
\left(\frac{m^{2}}{2e_{R}B}\right)
-2(2e_{R}B)^{2}\zeta_{H}'\left(-1,\frac{m^{2}}{(2e_{R}B)}\right)\right\}
. \eea

N\'{o}tese que, una vez renormalizada la carga, la densidad de energ\'{\i}a
de Casimir es independiente del par\'{a}metro de escala $\mu$.

\newpage

\section{Energ\'{i}a de vac\'{i}o en $3+1$ dimensiones con condiciones de contorno}
\label{periodico}

En esta secci\'{o}n, analizaremos el efecto de tama\~{n}o finito en la
direcci\'{o}n del campo sobre la energ\'{\i}a de vac\'{\i}o. Impondremos sobre
los campos de Dirac condiciones de contorno no locales, del
tipo\cite{twisted}

\be
\psi(r,\theta,L)=e^{i\alpha}\psi(r,\theta,0). \ee

Vemos que esta condici\'{o}n se reduce a la condici\'{o}n de periodicidad para
$\alpha=0$, mientras que de $\alpha=\pi$ se obtiene el problema con condiciones de
contorno antiperi\'{o}dicas.

Nuevamente  tenemos autovectores simult\'{a}neos de la energ\'{\i}a y del impulso en la
direcci\'{o}n del campo, s\'{o}lo que ahora el espectro de autovalores $p_{z}$ pasa a
ser discreto.

\be\begin{array}{ccccc}
  \textrm{Si} & e^{ip_{z}L}=e^{i\alpha} & \Rightarrow & p_{z}=\frac{2k\pi+\alpha}{L}
  & \textrm{donde $k$ es un entero}
\end{array}.
\ee

Entonces, en la expresi\'{o}n para la energ\'{\i}a de vac\'{\i}o, debemos ahora sumar
sobre el espectro discreto de autovalores $p_{z}$

\bea\label{E0L}
  E_{0}^{L}&=&-\frac{eBA}{2\pi}\sum_{k=-\infty}^{\infty}
  \left\{\left[m^{2}+\left(\frac{2\pi k + \alpha}{L}\right)^{2}\right]^{1/2}+
  2\sum_{n=1}^{\infty}  \left[m^{2}+\left(\frac{2\pi k+
  \alpha}{L}\right)^{2}+2eBn\right]^{1/2}\right\}\nn\\&&\nn\\
  &=&-\frac{eBA}{2\pi}\sum_{k=-\infty}^{\infty}\left\{2\sum_{n=0}^{\infty}
  \left[m^{2}+\left(\frac{2\pi k + \alpha}{L}\right)^{2}+2eBn\right]^{1/2}
  -\left[m^{2}+\left(\frac{2\pi k + \alpha}{L}\right)^{2}\right]^{1/2}\right\}  .
\eea

Una vez regularizada mediante la t\'{e}cnica de la funci\'{o}n zeta, (\ref{E0L}) se
escribe

\bea E_{0}^{L}&=&-\frac{eBA}{2\pi}\left(\mu^{s+1}\sum_{k=-\infty}^{\infty}
  \left\{2\sum_{n=0}^{\infty}\left[m^{2}+\left(\frac{2\pi k+
  \alpha}{L}\right)^{2}+2eBn\right]^{-s/2}\right.\right.-{}\nn\\ &&\nn\\
  &&\left.\left.{}-\left[m^{2}+\left(\frac{2\pi k+
  \alpha}{L}\right)^{2}\right]^{-s/2}\right\}\right)_{s=-1}.
\eea

Usamos la transformada de Mellin (\ref{mellin}) para luego poder hacer la suma en $n$

\bea E_{0}^{L}&=&-\frac{eBA}{2\pi}\left(\mu^{s+1}\sum_{k=-\infty}^{\infty}
  \left\{\sum_{n=0}^{\infty}\frac{2}{\Gamma(s/2)}\right.\right.\times{}\nn\\ &&\nn\\
  &&{}\times\int_{0}^{\infty}dt
  t^{s/2-1}\exp\left[-\left(m^{2}+\left(\frac{2\pi k+
  \alpha}{L}\right)^{2}+2eBn\right)t\right]-{}\nn\\
  &&\nn\\
&&\left.\left.{} -\frac{1}{\Gamma(s/2)}\int_{0}^{\infty}dt
  t^{s/2-1}\exp\left[-\left(m^{2}+\left(\frac{2\pi k+
  \alpha}{L}\right)^{2}\right)t\right]\right\}\right)_{s=-1}\nn\\
  &&\nn\\
&=&-\frac{eBA}{2\pi}\left(\mu^{s+1}\left\{\sum_{n=0}^{\infty}\frac{2}{\Gamma(s/2)}
\int_{0}^{\infty}dt
t^{s/2-1}\exp\left[-\left(m^{2}+\left(\frac{\alpha}{L}\right)^{2}+2eBn\right)t\right]
\right.\right.\times{}\nn\\ &&\nn\\ &&{}\times\sum_{k=-\infty}^{\infty}\exp
\left(-\frac{4\pi\alpha}{L^{2}}tk-\frac{4\pi^{2}}{L^{2}}tk^{2}\right)-{}\nn\\ &&\nn\\
&&{}- \frac{1}{\Gamma(s/2)} \int_{0}^{\infty}dt
t^{s/2-1}\exp\left[-\left(m^{2}+\left(\frac{\alpha}{L}\right)^{2}\right)t\right]\times{}\nn\\
&&\nn\\ &&\left.\left.{}\times\sum_{k=-\infty}^{\infty}\exp\left(-\frac{4\pi
\alpha}{L^{2}}tk-\frac{4\pi^{2}}{L^{2}}tk^{2}\right)\right\}\right)_{s=-1}. \eea

\bigskip

Podemos expresar este resultado en t\'{e}rminos de la funci\'{o}n theta de Jacobi, que
queda definida por \cite{rusa}

\be\label{theta} \theta(x,y)=\sum_{k=-\infty}^{\infty}e^{2\pi kx}e^{-\pi k^{2}y}. \ee

Por lo tanto

\be
\sum_{k=-\infty}^{\infty}\exp\left(-\frac{4\pi
\alpha}{L^{2}}tk-\frac{4\pi^{2}}{L^{2}}tk^{2}\right)=\theta\left(-\frac{2\alpha
t}{L^{2}},\frac{4\pi t}{L^{2}}\right) .\ee

Considerando la siguiente propiedad de inversi\'{o}n de la funci\'{o}n theta de Jacobi
\cite{ambj}

\be\label{the-pro} \theta(x,y)=\frac{1}{\sqrt{y}}\exp\left(\frac{\pi x^{2}}{y}\right)
\theta\left(\frac{x}{iy},\frac{1}{y}\right), \ee resulta

\bea \sum_{k=-\infty}^{\infty}\exp\left(-\frac{4\pi
\alpha}{L^{2}}tk-\frac{4\pi^{2}}{L^{2}}tk^{2}\right)&=&
\sqrt{\frac{\pi}{t}}\left(\frac{L}{2\pi}\right)\exp\left(\frac{\alpha^{2}t}{L^{2}}\right)
\theta\left(\frac{i\alpha}{2\pi},\frac{L^{2}}{4\pi t}\right)\nn\\ &&\nn\\
&=&\sqrt{\frac{\pi}{t}}\left(\frac{L}{2\pi}\right)\exp\left(\frac{\alpha^{2}t}{L^{2}}\right)
\sum_{k=-\infty}^{\infty}\exp\left(i\alpha k -\frac{L^{2}}{4t}k^{2}\right) .\eea

Reemplazando en la expresi\'{o}n para la energ\'{\i}a de vac\'{\i}o

\bea E_{0}^{L}&=&-\frac{eBA}{2\pi}\sqrt{\pi}\frac{L}{2\pi}\left(\mu^{s+1}
\left\{2\sum_{n=0}^{\infty}\frac{1}{\Gamma(s/2)}\int_{0}^{\infty}dt
t^{\frac{s-1}{2}-1}\exp\left[-\left(m^{2}+2eBn\right)t\right]\right.\right.\times{}\nn\\
&&\nn\\ &&{}\times\sum_{k=-\infty}^{\infty}
e^{ik\alpha}\exp\left(-\frac{L^{2}k^{2}}{4t}\right)-{} \nn\\ &&\nn\\
&&\left.\left.{}-\frac{1}{\Gamma(s/2)} \int_{0}^{\infty}dt
t^{\frac{s-1}{2}-1}\exp\left(-m^{2}t\right)\sum_{k=-\infty}^{\infty}e^{ik\alpha}
\exp\left(-\frac{L^{2}k^{2}}{4t}\right)\right\}\right)_{s=-1} .\eea

Nuevamente, calculamos la densidad de energ\'{\i}a de vac\'{\i}o por unidad de volumen.
Escribiendo por separado el t\'{e}rmino correspondiente a $k=0$ podemos identificar a
la densidad de energ\'{\i}a de vac\'{\i}o sin condiciones de contorno.

\bea
\varepsilon_{0}^{L}&=&\frac{eB}{(2\pi)^{2}}\left(\mu^{s+1}\frac{\sqrt{\pi}}{\Gamma(s/2)}
\left\{\int_{0}^{\infty}dt
t^{\frac{s-1}{2}-1}\exp(-m^{2}t)\right.\right.-{}\nn\\ &&\nn\\
&&\left.\left.{}-2\sum_{n=0}^{\infty}\int_{0}^{\infty}dt
t^{\frac{s-1}{2}-1}\exp\left[-\left(m^{2}+2eBn\right)t\right]\right\}\right)_{s=-1}+{}\nn\\
&&\nn\\
&&{}+\frac{eB}{(2\pi)^{2}}\left(\mu^{s+1}\frac{\sqrt{\pi}}{\Gamma(s/2)}
\left\{\int_{0}^{\infty}dt
t^{\frac{s-1}{2}-1}\exp(-m^{2}t)2\sum_{k=1}^{\infty}e^{ik\alpha}
\exp\left(-\frac{L^{2}k^{2}}{4t}\right)\right.\right.-{}\nn\\
&&\nn\\ &&\left.\left.{}-2\sum_{n=0}^{\infty}\int_{0}^{\infty}dt
t^{\frac{s-1}{2}-1}\exp\left[-\left(m^{2}+2eBn\right)t\right]2\sum_{k=1}^{\infty}e^{ik\alpha}
\exp\left(-\frac{L^{2}k^{2}}{4t}\right)\right\}\right)_{s=-1}.\eea

Efectivamente, los dos primeros t\'{e}rminos corresponden a la densidad de energ\'{\i}a
de vac\'{\i}o $\varepsilon_{0}$ del problema en todo el espacio estudiado en la
secci\'{o}n previa (ver ecuaci\'{o}n (\ref{E0cita}))

\bea\varepsilon_{0}^{L}&=&\varepsilon_{0}
-\frac{2eB}{(2\pi)^{2}}\left\{\frac{\sqrt{\pi}\mu^{s+1}}{\Gamma(s/2)}
\sum_{k=1}^{\infty}e^{ik\alpha}\int_{0}^{\infty}dt
t^{\frac{s-1}{2}-1}\exp(-m^{2}t)\right.\times{}\nn\\ &&\nn\\
&&\left.{}\times\left[2\sum_{n=0}^{\infty}(e^{-2eBt})^{n}-1\right]
\exp\left(-\frac{L^{2}k^{2}}{4t}\right)\right\}_{s=-1}. \eea

A fin de estudiar la variaci\'{o}n de energ\'{\i}a debida al tama\~{n}o finito,
realizamos la suma geom\'{e}trica, cuyo resultado es

\begin{equation}\sum_{n=0}^{\infty}x^{n}=\frac{1}{1-x}\quad\textrm{para}\quad x<1,
\end{equation}
y evaluamos para $s=-1$

\be\label{E0Lint}\varepsilon_{0}^{L}=\varepsilon_{0}+\frac{eB}{(2\pi)^{2}}
\sum_{k=1}^{\infty}e^{ik\alpha}\int_{0}^{\infty}dt
t^{-2}\exp(-m^{2}t)
\coth(eBt)\exp\left(-\frac{L^{2}k^{2}}{4t}\right).\ee

Cabe se\~{n}alar aqu\'{\i} que, para $\alpha=0,\,\pi$, este resultado coincide con el
encontrado, mediante el esquema de regularizaci\'{o}n del tiempo propio en
\cite{cougo}.

A fin de obtener una expresi\'{o}n m\'{a}s expl\'{\i}cita para la densidad de
energ\'{\i}a de vac\'{\i}o, usamos el siguiente desarrollo para la cotangente
hiperb\'{o}lica

\be\label{coth} x\coth(x)=1+\frac{x^{2}}{3}-2x^{4}
\sum_{q=1}^{\infty}\frac{1}{\pi^{2}q^{2}(x^{2}+\pi^{2}q^{2})}.\ee

\medskip

La contribuci\'{o}n a (\ref{E0Lint}) del primer t\'{e}rmino del desarrollo es

\bea &=&\frac{1}{(2\pi)^{2}} \sum_{k=1}^{\infty}e^{ik\alpha}\int_{0}^{\infty}dt
t^{-3}\exp\left(-m^{2}t-\frac{L^{2}k^{2}}{4t}\right)\nn\\ &&\nn\\
&=&\frac{1}{(2\pi)^{2}} \sum_{k=1}^{\infty}e^{ik\alpha}\int_{0}^{\infty}dx
x\exp\left(-\frac{m^{2}}{x}-\frac{L^{2}k^{2}x}{4}\right) .\eea

La integral puede resolverse con el siguiente resultado
\cite{rusa}, el cual es v\'{a}lido para $Re(\beta)>0$ y $Re(\gamma)>0$

\be\label{be-mod} \int_{0}^{\infty}dx x^{\nu-1}\exp\left(-\frac{\beta}{x}-\gamma x
\right)=2\left(\frac{\beta}{\gamma}\right)^{\nu/2}K_{\nu}\left(2\sqrt{\beta\gamma}\right),\ee
donde $K_{\nu}$ son las funciones de Bessel modificadas. La contribuci\'{o}n de este
t\'{e}rmino, independiente del campo, es

\be\label{K2}
\varepsilon_{0}^{c}=\frac{2m^{2}}{\pi^{2}L^{2}}
\sum_{k=1}^{\infty}\frac{e^{ik\alpha}}{k^{2}}K_{2}(mLk).\ee

Este t\'{e}rmino se reduce, para $\alpha=0$, al resultado bien conocido para la
energ\'{\i}a de Casimir de un campo de Dirac en una caja con condiciones peri\'{o}dicas
\cite{ambj}.

\medskip

El segundo t\'{e}rmino del desarrollo (\ref{coth}) dar\'{a} una contribuci\'{o}n
cuadr\'{a}tica en los campos

\bea\label{K0}
&=&\frac{1}{3}\frac{(eB)^{2}}{(2\pi)^{2}}\sum_{k=1}^{\infty}e^{ik\alpha}\int_{0}^{\infty}dt
t^{-1}\exp\left(-m^{2}t-\frac{L^{2}k^{2}}{4t}\right)\nn\\ &&\nn\\
&=&\frac{2}{3}\left(\frac{eB}{2\pi}\right)^{2}\sum_{k=1}^{\infty}e^{ik\alpha}K_{0}(mLk).
\eea donde nuevamente hemos usado (\ref{be-mod}) para resolver la integral.

Podemos analizar la convergencia  de estos dos primeros t\'{e}rminos a partir del
comportamiento asint\'{o}tico de las funciones de Bessel modificadas para argumentos
grandes \cite{rusa}. Para $z\rightarrow\infty$

\be K_{\nu}(z)=\sqrt{\frac{\pi}{2z}}e^{-z}\left[1+O\left(\frac{1}{z}\right)\right],\ee
independientemente de $\nu$. Vemos que el factor $e^{-z}$ garantiza la convergencia de
las sumas (\ref{K2}) y (\ref{K0})

\bigskip

Para evaluar el \'{u}ltimo t\'{e}rmino en (\ref{coth}) sumamos y restamos
el t\'{e}rmino que corresponder\'{\i}a a $k=0$ para poder extender la suma
desde $-\infty$ hasta $+\infty$ y poder reconstruir la funci\'{o}n
theta de Euler.

\bea &=&-2\frac{(eB)^{4}}{(2\pi)^{2}}\sum_{k=1}^{\infty}e^{ik\alpha}
\int_{0}^{\infty}dt\cdot t \sum_{q=1}^{\infty}
\frac{\exp\left(-m^{2}t-\frac{L^{2}k^{2}}{4t}\right)}{\pi^{2}q^{2}[(eBt)^{2}+\pi^{2}q^{2}]}\nn\\
&&\nn\\ &=&-\frac{(eB)^{4}}{(2\pi)^{2}}\sum_{k=-\infty}^{\infty}e^{ik\alpha}
\int_{0}^{\infty}dt\cdot t \sum_{q=1}^{\infty}
\frac{\exp\left(-m^{2}t-\frac{L^{2}k^{2}}{4t}\right)}{\pi^{2}q^{2}[(eBt)^{2}+\pi^{2}q^{2}]}+{}\nn\\
&&\nn\\ &&{}+\frac{(eB)^{4}}{(2\pi)^{2}} \int_{0}^{\infty}dt\cdot t \sum_{q=1}^{\infty}
\frac{\exp(-m^{2}t)}{\pi^{2}q^{2}[(eBt)^{2}+\pi^{2}q^{2}]}.\eea

En el primer t\'{e}rmino podemos formar el siguiente factor y
reescribirlo usando la definici\'{o}n de la funci\'{o}n $\theta(x,y)$ y su
propiedad de inversi\'{o}n (\ref{the-pro})

\be
\sum_{k=-\infty}^{\infty}e^{ik\alpha}e^{-\frac{L^{2}k^{2}}{4t}}
=\sqrt{\frac{t}{\pi}}\left(\frac{2\pi}{L}\right)\sum_{k=-\infty}^{\infty}
\exp\left[-\left(\frac{2k\pi+\alpha}{L}\right)^{2}t\right]. \ee

\hfill\break

La contribuci\'{o}n a la densidad de energ\'{\i}a de vac\'{\i}o es

\bea
&=&-\frac{(eB)^{2}}{2\pi^{3/2}L}\sum_{k=-\infty}^{\infty}\sum_{q=1}^{\infty}\frac{1}{\pi^{2}q^{2}}
\int_{0}^{\infty}dt.t^{3/2}
\frac{\exp\left[-m^{2}t-\left(\frac{2k\pi+\alpha}{L}\right)^{2}t\right]}
{\left[t^{2}+\left(\frac{q\pi}{eB}\right)^{2}\right]}+{}\nn\\ &&\nn\\ &&{}+
\left(\frac{eB}{2\pi}\right)^{2}\sum_{q=1}^{\infty}\frac{1}{\pi^{2}q^{2}}
\int_{0}^{\infty}dt.t\frac{\exp(-m^{2}t)}{\left[t^{2}+\left(\frac{q\pi}{eB}\right)^{2}\right]}.
\eea

Las integrales las resolvemos usando la siguiente soluci\'{o}n que es
v\'{a}lida para $Re(\beta)>0$, $Re(\mu)>0$ y $Re(\nu)>-1$ \cite{rusa}.

\bea \int_{0}^{\infty}\frac{x^{\nu}\exp(-\mu x)}{x^{2}+\beta^{2}}&=&
\frac{\Gamma(\nu)}{2}\beta^{\nu-1}\left[\exp\left(i\mu\beta+i\frac{(\nu-1)\pi}{2}\right)
\Gamma(1-\nu,i\mu\beta)\right.+{}\nn\\ &&\nn\\
&&\left.{}+\exp\left(-i\mu\beta-i\frac{(\nu-1)\pi}{2}\right)
\Gamma(1-\nu,-i\mu\beta)\right], \eea donde $\Gamma(a,z)$ es la funci\'{o}n gamma
incompleta \cite{Abram}.

\bigskip

La contribuci\'{o}n a la densidad de energ\'{\i}a es

\bea
&-&\frac{(eB)^{3/2}}{8\pi^{5/2}L}\sum_{k=-\infty}^{\infty}\sum_{q=1}^{\infty}\frac{1}{q^{3/2}}
\left\{\exp\left(\frac{iq\pi}{eB}\left[m^{2}+\left(\frac{2k\pi+\alpha}{L}\right)^{2}\right]
+\frac{i\pi}{4}\right)\right.\times{}\nn\\ &&\nn\\ &&
{}\times\Gamma\left(-1/2,\frac{iq\pi}{eB}
\left[m^{2}+\left(\frac{2k\pi+\alpha}{L}\right)^{2}\right]\right)+{}\nn\\ &&\nn\\
&+&\left.\exp\left(-\frac{iq\pi}{eB}\left[m^{2}+\left(\frac{2k\pi+\alpha}{L}\right)^{2}\right]
-\frac{i\pi}{4}\right)\Gamma\left(-1/2,-\frac{iq\pi}{eB}
\left[m^{2}+\left(\frac{2k\pi+\alpha}{L}\right)^{2}\right]\right)\right\}+{}\nn\\
&&\nn\\ &+&\frac{(eB)^{2}}{8\pi^{4}}\sum_{q=1}^{\infty}\frac{1}{q^{2}}
\left[\exp\left(\frac{im^{2}q\pi}{eB}\right)\Gamma\left(0,\frac{im^{2}q\pi}{eB}
\right)+\exp\left(-\frac{im^{2}q\pi}{eB}\right)\Gamma\left(0,-\frac{im^{2}q\pi}{eB}
\right)\right] .\eea

\bigskip

Para verificar la convergencia de las series, consideramos el comportamiento
asint\'{o}tico de las funciones gamma incompletas \cite{rusa}. Para
$z\rightarrow\infty$ en $|\arg(z)|<\frac{3\pi}{2}$.

\be
\Gamma(a,z)\sim z^{a-1}e^{-z}\left[1+\frac{a-1}{z}+\frac{(a-1)(a-2)}{z^{2}}+...\right].
\ee

Para $k$ y/o $q$ suficientemente grandes, los sumandos se comportan de la siguiente
manera: en ambos casos los primeros t\'{e}rminos del comportamiento asint\'{o}tico se
cancelan entre los dos t\'{e}rminos de cada sumando. Para el sumando de la primera suma
(doble en este caso)

\be
\sim -\frac{3}{q^{4}}.\frac{e^{i\pi/4}(eB)^{5/2}}{\left\{i\pi
\left[m^{2}+\left(\frac{2k\pi+\alpha}{L}\right)^{2}\right]\right\}^{5/2}}. \ee

Para el segundo sumando, que s\'{o}lo se suma en $q$

\be
\sim -\frac{(eB)^{2}}{\pi^{2}m^{2}q^{4}} .\ee

Concluimos que las sumas de estos dos \'{u}ltimos t\'{e}rminos de la densidad de
energ\'{\i}a de vac\'{\i}o son convergentes. La expresi\'{o}n que finalmente damos para
la densidad de energ\'{\i}a de vac\'{\i}o del campo de fermiones en interacci\'{o}n con
un campo magn\'{e}tico uniforme y con una condici\'{o}n de contorno en la direcci\'{o}n
del campo es la siguiente

\bea \label{E0Lfinal}\varepsilon_{0}^{L}&=&\varepsilon_{0}+\varepsilon_{0}^{c}+
\frac{2}{3}\left(\frac{eB}{2\pi}\right)^{2}\sum_{k=1}^{\infty}e^{ik\alpha}K_{0}(mLk)-{}\nn\\&&\nn\\
&&{}-\frac{(eB)^{3/2}}{8\pi^{5/2}L}\sum_{k=-\infty}^{\infty}\sum_{q=1}^{\infty}\frac{1}{q^{3/2}}
\left\{\exp\left(\frac{iq\pi}{eB}\left[m^{2}+\left(\frac{2k\pi+\alpha}{L}\right)^{2}\right]
+\frac{i\pi}{4}\right)\right.\times{}\nn\\
&&\nn\\&&{}{}\times\Gamma\left(-1/2,\frac{iq\pi}{eB}
\left[m^{2}+\left(\frac{2k\pi+\alpha}{L}\right)^{2}\right]\right)+{}\nn\\ &&\nn\\
&&\left.{}+\exp\left(-\frac{iq\pi}{eB}\left[m^{2}+\left(\frac{2k\pi+\alpha}{L}\right)^{2}\right]
-\frac{i\pi}{4}\right)\Gamma\left(-1/2,-\frac{iq\pi}{eB}
\left[m^{2}+\left(\frac{2k\pi+\alpha}{L}\right)^{2}\right]\right)\right\}+{}\nn\\
&&\nn\\ &&{}+\frac{(eB)^{2}}{8\pi^{4}}\sum_{q=1}^{\infty}\frac{1}{q^{2}}
\left[\exp\left(\frac{im^{2}q\pi}{eB}\right)\Gamma\left(0,\frac{im^{2}q\pi}{eB}
\right)+\exp\left(-\frac{im^{2}q\pi}{eB}\right)\Gamma\left(0,-\frac{im^{2}q\pi}{eB}
\right)\right]. \eea

En esta expresi\'{o}n de la densidad de energ\'{\i}a de vac\'{\i}o hemos aislado una
contribuci\'{o}n, $\varepsilon_{0}$, independiente de las condiciones de contorno y
otra, $\varepsilon_{0}^{c}$, debida exclusivamente al tama\~{n}o finito e independiente
del campo. Si definimos la densidad de energ\'{\i}a de Casimir como la diferencia entre
$\varepsilon_{0}^{L}$ en (\ref{E0Lfinal}) y su valor para $B=0$ y
$L\rightarrow\infty$,obtenemos (usando $\varepsilon_{0}$ de (\ref{E0polo}))

\bea
\varepsilon_{Cas}&=&\varepsilon_{0}^{L}-\varepsilon_{0}^{l}\nn\\
&&\nn\\
&=&\left[\frac{\frac{(2eB)^{2}}{6}}{8\pi^{2}(s+1)}\right]_{s=-1}
+\frac{1}{16\pi^{2}}\left\{\left[\psi(1)-\psi(-1/2)+1+\ln\left(\frac{m^{2}}{2eB}\right)
-\ln\left(\frac{m^{2}}{\mu^{2}}\right)\right]\frac{(2eB)^{2}}{6}\right.+{}\nn\\&&\nn\\
&&\left.{}+m^{4}\left[\ln\left(\frac{m^{2}}{2eB}\right)-\frac{1}{2}\right]
-m^{2}2eB\ln\left(\frac{m^{2}}{2eB}\right)-
2(2eB)^{2}\zeta_{H}'\left(-1,\frac{m^{2}}{2eB}\right)\right\}+{}\nn\\
&&\nn\\ &&{}+\varepsilon_{0}^{c}+
\frac{2}{3}\left(\frac{eB}{2\pi}\right)^{2}\sum_{k=1}^{\infty}e^{ik\alpha}K_{0}(mLk)-{}\nn\\&&\nn\\
&&{}-\frac{(eB)^{3/2}}{8\pi^{5/2}L}\sum_{k=-\infty}^{\infty}\sum_{q=1}^{\infty}\frac{1}{q^{3/2}}
\left\{\exp\left(\frac{iq\pi}{eB}\left[m^{2}+\left(\frac{2k\pi+\alpha}{L}\right)^{2}\right]
+\frac{i\pi}{4}\right)\right.\times{}\nn\\
&&\nn\\&&{}{}\times\Gamma\left(-1/2,\frac{iq\pi}{eB}
\left[m^{2}+\left(\frac{2k\pi+\alpha}{L}\right)^{2}\right]\right)+{}\nn\\
&&\nn\\
&&\left.{}+\exp\left(-\frac{iq\pi}{eB}\left[m^{2}+\left(\frac{2k\pi+\alpha}{L}\right)^{2}\right]
-\frac{i\pi}{4}\right)\Gamma\left(-1/2,-\frac{iq\pi}{eB}
\left[m^{2}+\left(\frac{2k\pi+\alpha}{L}\right)^{2}\right]\right)\right\}+{}\nn\\
&&\nn\\
&&{}+\frac{(eB)^{2}}{8\pi^{4}}\sum_{q=1}^{\infty}\frac{1}{q^{2}}
\left[\exp\left(\frac{im^{2}q\pi}{eB}\right)\Gamma\left(0,\frac{im^{2}q\pi}{eB}
\right)+\exp\left(-\frac{im^{2}q\pi}{eB}\right)\Gamma\left(0,-\frac{im^{2}q\pi}{eB}
\right)\right]. \eea

\bigskip

Nuevamente, los t\'{e}rminos proporcionales a $B^{2}$ pueden eliminarse por
renormalizaci\'{o}n de la carga el\'{e}ctrica. En este caso, debido al tama\~{n}o
finito, aparecer\'{a}n nuevas contribuciones a la carga renormalizada (por ejemplo, el
tercer t\'{e}rmino en (\ref{E0Lfinal})). Para ver que, en efecto, \'{e}ste es el
\'{u}nico t\'{e}rmino que agrega una nueva contribuci\'{o}n a la carga renormalizada,
la calculamos expl\'{\i}citamente a partir de su definici\'{o}n (\ref{eR}) y de
(\ref{E0Lint}). Para ello, debemos mostrar que la serie derivada en (\ref{E0Lint}) es
uniformemente convergente. El segundo t\'{e}rmino en (\ref{E0Lint}) se escribe, al
desarrollar la cotangente hiperb\'{o}lica,

\begin{equation}
\frac{1}{(2\pi)^{2}} \sum_{k=1}^{\infty}e^{ik\alpha}\int_{0}^{\infty}dt t^{-3}
\left(1+2\sum_{q=1}^{\infty}\frac{(eBt)^{2}}{(eBt)^{2}+\pi^{2}q^{2}}\right)
\exp\left(-m^{2}t-\frac{L^{2}k^{2}}{4t}\right).
\end{equation}

\hfill\break

La serie derivada resulta

\begin{equation}
\frac{2e^{2}}{(2\pi)^{2}} \sum_{k=1}^{\infty}e^{ik\alpha}\int_{0}^{\infty}dt t^{-1}
\sum_{q=1}^{\infty}\frac{\pi^{2}q^{2}}{[(eBt)^{2}+\pi^{2}q^{2}]^{2}}
\exp\left(-m^{2}t-\frac{L^{2}k^{2}}{4t}\right),
\end{equation}y el m\'{o}dulo de sus t\'{e}rminos puede acotarse con los t\'{e}rminos
independientes de $B$ de una serie convergente

\bea &&\left|\frac{2e^{2}}{(2\pi)^{2}} e^{ik\alpha}\int_{0}^{\infty}dt t^{-1}
\sum_{q=1}^{\infty}\frac{\pi^{2}q^{2}}{[(eBt)^{2}+\pi^{2}q^{2}]^{2}}
\exp\left(-m^{2}t-\frac{L^{2}k^{2}}{4t}\right)\right|\leq{}\nn\\ &&\nn\\
&&{}\leq\frac{2e^{2}}{(2\pi)^{2}}\frac{\zeta_{R}(2)}{\pi^{2}} \int_{0}^{\infty}dt
t^{-1}\exp\left(-m^{2}t-\frac{L^{2}k^{2}}{4t}\right)
=\frac{e^{2}}{\pi^{4}}\zeta_{R}(2)K_{0}(mLk).\eea

\bigskip

Concluimos que la serie derivada en (\ref{E0Lint}) es uniformemente convergente y, por
lo tanto, podemos evaluar su derivada t\'{e}rmino a t\'{e}rmino. As\'{\i}, obtenemos

\bea &&\frac{\partial}{\partial B^{2}}\left[\frac{1}{(2\pi)^{2}}
\sum_{k=1}^{\infty}e^{ik\alpha}\int_{0}^{\infty}dt t^{-3}
\left(1+2\sum_{q=1}^{\infty}\frac{(eBt)^{2}}{(eBt)^{2}+\pi^{2}q^{2}}\right)
\exp\left(-m^{2}t-\frac{L^{2}k^{2}}{4t}\right)\right]_{B^{2}=0}={}\nn\\ &&\nn\\
&&{}=\frac{e^{2}}{6\pi^{2}}\sum_{k=1}^{\infty}e^{ik\alpha}K_{0}(mLk). \eea

\bigskip

De esta manera, se agrega un t\'{e}rmino en la carga renormalizada  que depende del
tama\~{n}o finito en la direcci\'{o}n del campo

\bea
  \frac{\mu^{4}}{8 e_{R}^{2}}&=&\frac{\mu^{4}}{8 e^{2}}+\frac{\mu^{4}}{96\pi^{2}}
  \left[\frac{2}{s+1}+\psi(1)-\psi(-1/2)-\ln c\right]_{s=-1}+{}\nn\\ &&\nn\\
  &&{}+\frac{\mu^{4}}{24\pi^{2}}\sum_{k=1}^{\infty}e^{ik\alpha}K_{0}(mLk). \eea

\hfill\break

\hfill\break

\hfill\break

La expresi\'{o}n que finalmente nos queda para la densidad de energ\'{\i}a de Casimir

\bea
  \varepsilon_{Cas}&=& \frac{B^{2}}{2}
+\frac{1}{16\pi^{2}}\left\{\frac{(2e_{R}B)^{2}}{6}\left[\ln
\left(\frac{m^{2}}{2e_{R}B}\right) +1\right]
+m^{4}\left[\ln\left(\frac{m^{2}}{2e_{R}B}\right)-\frac{1}{2}\right]\right.-{}\nn\\
&&\nn\\ &&\left.{}-m^{2}2e_{R}B \ln \left(\frac{m^{2}}{2e_{R}B}\right)
-2(2e_{R}B)^{2}\zeta_{H}'\left(-1,\frac{m^{2}}{(2e_{R}B)}\right)\right\}+\varepsilon_{0}^{c}
-{}\nn\\ &&\nn\\&&{}
-\frac{(e_{R}B)^{3/2}}{8\pi^{5/2}L}\sum_{k=-\infty}^{\infty}\sum_{q=1}^{\infty}\frac{1}{q^{3/2}}
\left\{\exp\left(\frac{iq\pi}{e_{R}B}\left[m^{2}+\left(\frac{2k\pi+\alpha}{L}\right)^{2}\right]
+\frac{i\pi}{4}\right)\right.\times{}\nn\\
&&\nn\\&&{}{}\times\Gamma\left(-1/2,\frac{iq\pi}{e_{R}B}
\left[m^{2}+\left(\frac{2k\pi+\alpha}{L}\right)^{2}\right]\right)+{}\nn\\ &&\nn\\
&&\left.{}+\exp\left(-\frac{iq\pi}{e_{R}B}\left[m^{2}+\left(\frac{2k\pi+\alpha}{L}\right)^{2}\right]
-\frac{i\pi}{4}\right)\Gamma\left(-1/2,-\frac{iq\pi}{e_{R}B}
\left[m^{2}+\left(\frac{2k\pi+\alpha}{L}\right)^{2}\right]\right)\right\}+{}\nn\\
&&\nn\\ &&{}+\frac{(e_{R}B)^{2}}{8\pi^{4}}\sum_{q=1}^{\infty}\frac{1}{q^{2}}
\left[\exp\left(\frac{im^{2}q\pi}{e_{R}B}\right)\Gamma\left(0,\frac{im^{2}q\pi}{e_{R}B}
\right)+\exp\left(-\frac{im^{2}q\pi}{e_{R}B}\right)\Gamma\left(0,-\frac{im^{2}q\pi}{e_{R}B}
\right)\right] . \eea

Otra vez debe destacarse que, una vez renormalizada la carga, la densidad de
energ\'{\i}a de Casimir no depende del par\'{a}metro de escala $\mu$.

\newpage

\section{Resumen de los resultados obtenidos}

En este trabajo, hemos calculado las correcciones cu\'{a}nticas a la energ\'{\i}a de
vac\'{\i}o (e\-ner\-g\'{\i}\-as de Casimir) para campos de Dirac en presencia de un
campo magn\'{e}tico uniforme de background $\vec{B}$. Hemos analizado las correcciones
debidas exclusivamente a la presencia del campo de background y al efecto combinado de
dicha presencia y la imposici\'{o}n de condiciones de contorno. En todos los casos
hemos utilizado el m\'{e}todo de suma de modos, combinado con la regularizaci\'{o}n
zeta \cite{Dowker:1976tf,zeta}.

\medskip

En primer lugar, hemos evaluado la energ\'{\i}a de Casimir para campos de Dirac
confinados en un plano perpendicular a la direcci\'{o}n de $\vec{B}$ (problema en $2+1$
dimensiones). En este caso, la energ\'{\i}a de Casimir, definida por extensi\'{o}n
anal\'{\i}tica, resulta finita y no requiere, por lo tanto, una renormalizaci\'{o}n de
los par\'{a}metros de la teor\'{\i}a.

\medskip

A continuaci\'{o}n, hemos tratado el problema en $3+1$ dimensiones, cuando todo el
espacio tridimensional es accesible a los fermiones. En este caso, la correcci\'{o}n
cu\'{a}ntica a la energ\'{\i}a de vac\'{\i}o resulta divergente, con residuos en los
polos de dos tipos: independientes de $B$ y proporcionales a $B^{2}$. La primera de
estas divergencias desaparece en el c\'{a}lculo de la energ\'{\i}a de Casimir, dado que
est\'{a} presente tambi\'{e}n en ausencia del campo magn\'{e}tico. Los t\'{e}rminos
proporcionales a $B^{2}$ pueden eliminarse por renormalizaci\'{o}n de la carga
el\'{e}ctrica.

\medskip

Finalmente, hemos tratado el efecto combinado de campo de background y condiciones de
contorno no locales ("twisted")\cite{twisted}, cuando el espacio accesible a los
fermiones tiene tama\~{n}o finito en la direcci\'{o}n de $\vec{B}$. En esta
situaci\'{o}n, hemos podido separar las contribuciones a la energ\'{\i}a de vac\'{\i}o
debidas a uno y otro factor externo exclusivamente de aquella debida a la acci\'{o}n
combinada de ambos. Aqu\'{\i}, nuevamente, debimos renormalizar la carga el\'{e}ctrica
a fin de obtener un resultado finito para la energ\'{\i}a de Casimir

\medskip

En los casos sin borde, hemos verificado la consistencia con
resultados previos de otros autores \cite{blauviss} para la acci\'{o}n
efectiva (ya que para campo el\'{e}ctrico nulo debe coincidir con la
energ\'{\i}a de vac\'{\i}o). En el caso con condiciones "twisted" hemos
mostrado que, en los l\'{\i}mites particulares de condiciones
peri\'{o}dicas y antiperi\'{o}dicas, nuestra energ\'{\i}a de vac\'{\i}o, expresada
en forma integral, coincide con resultados anteriores
\cite{cougo}. En este trabajo, hemos encontrado, para un par\'{a}metro
de "periodicidad" arbitrario, una expresi\'{o}n m\'{a}s expl\'{\i}cita de la
energ\'{\i}a de Casimir, como serie de funciones especiales.

\bigskip

\bigskip

\bigskip

\textbf{Agradecimientos:} Quisiera expresar mi agradecimiento a Horacio Falomir, a
Gabriela Beneventano, a Pablo Pisani y a Karin R\'{e}vora por sus valiosos comentarios
y sugerencias. Tambi\'{e}n quisiera agradecer a la Fundaci\'{o}n Antorchas por el apoyo
recibido, fundamental para la realizaci\'{o}n de este trabajo. Por \'{u}ltimo,
agradezco a Mariel Santangelo la confianza depositada en m\'{\i}, su inestimable
dedicaci\'{o}n y su gu\'{\i}a durante estos \'{u}ltimos dos a\~{n}os.

\newpage

\section*{Ap\'{e}ndice A: Soluciones de la ecuaci\'{o}n de Kummer}
\label{kummer}

Las soluciones independientes de la ecuaci\'{o}n de Kummer

\begin{equation}
  z\frac{d^{2}w}{dz^{2}}+(b-z)\frac{dw}{dz}-aw=0
\end{equation}
se pueden expresar en t\'{e}rminos de

\begin{equation}
  M(a,b,z)=1+\frac{az}{b}+\frac{a(a+1)z^{2}}{b(b+1)2!}+...
  +\frac{a(a+1)...(a+n-1)z^{n}}{b(b+1)...(b+n-1)n!}+...,
\end{equation}
definida s\'{o}lo para $b$ distinto de un entero negativo o cero y

\begin{equation}
  U(a,b,z)=\frac{\pi}{\sin \pi b}\left[\frac{M(a,b,z)}{\Gamma(1+a-b)\Gamma(b)}
  -z^{1-b}\frac{M(1+a-b,2-b,z)}{\Gamma(a)\Gamma(2-b)}\right],
\end{equation}
si bien estas dos funciones son soluciones de la ecuaci\'{o}n de
Kummer, no siempre resultan linealmente independientes.
\bigskip

Para estudiar los distintos casos consideraremos las cuatro siguientes soluciones

\be
\begin{array}{l}
  y_{1}(a,b,z)=M(a,b,z) \\
  y_{2}(a,b,z)=z^{1-b}M(1+a-b,2-b,z) \\
  y_{3}(a,b,z)=U(a,b,z) \\
  y_{4}(a,b,z)=e^{z}U(b-a,b,-z)
\end{array}\ee

Estudiaremos los comportamientos asint\'{o}ticos de las soluciones en los distintos
casos, para $z\rightarrow \infty$ y para $z\rightarrow 0$, y veremos para que valores
de $a$ y $b$ conducen a autofunciones de cuadrado integrable, recordando que en nuestro
problema $b$ s\'{o}lo toma valores enteros. En particular consideraremos el
comportamiento de $M(a,b,z)$ para $z$ grande, cuando $a$ es distinto de un entero
negativo,

\begin{equation}
  M(a,b,z)=\frac{\Gamma(b)}{\Gamma(a)}e^{z}z^{a-b}\left[1+O(|z|^{-1})\right]
  \quad\textrm{para}\quad z\rightarrow \infty,\quad Re(z)>0
\end{equation}
y el comportamiento de  $U(a,b,z)$ en un entorno del origen

\begin{equation}
  U(a,b,z)=\left\{\begin{array}{cc}
    \frac{\Gamma(b-1)}{\Gamma(a)}z^{1-b}+ O(|z|^{b-2})& \textrm{si } b>2 \\
     & \\
    \frac{1}{\Gamma(a)}z^{-1}+O(|\ln z|) & \textrm{si } b=2  \\
     & \\
    -\frac{1}{\Gamma(a)}[\ln z +\psi(a)]+ O(|z\ln z|)& \textrm{si } b=1  \\
     & \\
    \frac{1}{\Gamma(1+a)}+ O(|z\ln z|)& \textrm{si } b=0  \\
     & \\
     \frac{\Gamma(1-b)}{\Gamma(1+a-b)}+O(|z|) & \textrm{si } b<0 \
  \end{array}\right.
\end{equation}

\bigskip
\begin{enumerate}
\item  \emph{$a\neq -n$ (negativo o cero) y $b$ entero $b\geq 1$}

La soluci\'{o}n general es

\begin{equation}\
w(z)=A y_{1}(a,b,z)+B y_{3}(a,b,z),
\end{equation}
ya que su wronskiano es no nulo

\begin{equation}\
W(y_{1},y_{3})=-\frac{\Gamma(b)}{\Gamma(a)}z^{-b}e^{z}
\end{equation}

Si $z\rightarrow \infty$, entonces $y_{1}\sim e^{z}z^{a-b}$ y, por
lo tanto, las autofunciones que se obtienen a partir de esta
soluci\'{o}n divergen exponencialmente.

Si $z\rightarrow 0,\quad y_{3}\sim z^{1-b}$ para $b\geq 2$ y
entonces las autofunciones, que divergen como $\rho^{-k}$ para
$k\geq 1$ en el origen, no son de cuadrado integrable. Mientras
que para $b=1$ la divergencia es logar\'{\i}tmica y en consecuencia
esta componente del espinor s\'{\i} es de cuadrado integrable. Sin
embargo, al derivar la componente restante de (\ref{fg2}) la
divergencia en el origen va como $k^{-1}$.

\medskip

\item \emph{$a\neq -n$ (negativo o cero) y $b$ entero $b\leq 0$}

En este caso, la soluci\'{o}n general es

\begin{equation}\
w(z)=A y_{2}(a,b,z)+B y_{3}(a,b,z),
\end{equation}
dado que son soluciones linealmente independientes

\begin{equation}\
W(y_{2},y_{3})=-\frac{\Gamma(2-b)}{\Gamma(1+a-b)}z^{-b}e^{z}
\end{equation}

Si $z\rightarrow \infty$, entonces $y_{2}\sim e^{z}z^{a-b}$ y ,por
lo tanto, las autofunciones que se obtienen a partir de esta
soluci\'{o}n divergen exponencialmente.

Si $z\rightarrow 0,\quad y_{2}\sim 1$ y entonces las autofunciones
divergen como $\rho^{k}$ con $k\leq -1$ en el origen y no son de
cuadrado integrable.

\medskip

\item \emph{$a=-n$ (negativo o cero) y $b$ entero $b\geq 1$}

Escribimos la soluci\'{o}n general como

\begin{equation}\
w(z)=A y_{1}(a,b,z)+B y_{4}(a,b,z),
\end{equation}
para las cuales

\begin{equation}\
W(y_{1},y_{4})=\frac{\Gamma(b)}{\Gamma(b-a)}e^{i\epsilon\pi b}z^{-b}e^{z}
\end{equation}

Entonces $y_{1}=M(-n,b,z)=\frac{n!}{(b+1)(b+2)...(b-1+n)} L_{n}^{b-1}$, polinomio de
Laguerre generalizado que conduce a autofunciones de cuadrado integrable.

Mientras que $y_{4}=e^{z}U(b+n,b,-z)\sim e^{z}(-z)^{-b-n}$ diverge exponencialmente,
para $z\rightarrow\infty$.

\medskip

\item \emph{$a=-n$ (negativo o cero) y $b$ entero $-n+1\leq b\leq 0$}

  La soluci\'{o}n general es ahora

\begin{equation}\
w(z)=A y_{2}(a,b,z)+B y_{4}(a,b,z),
\end{equation}
y el wronskiano de estas soluciones es

\begin{equation}\
W(y_{2},y_{4})=-\frac{\Gamma(2-b)}{\Gamma(1-a)}z^{-b}e^{z}
\end{equation}

Tenemos
$y_{2}=z^{1-b}M(-n+1-b,2-b,z)=z^{1-b}\frac{(n+b-1)!}{(2-b)(2-b+1)...(2-b+n-1)}L_{n+b-1}^{-b+1}$,
con las que resultan autofunciones de cuadrado integrable.

Por el contrario, las soluciones $y_{4}=e^{z}U(b+n,b,-z)\sim e^{z}(-z)^{-b-n}$ divergen
exponencialmente,para $z\rightarrow\infty$.

\medskip

\item \emph{$a=-n$ (negativo o cero) y $b$ entero $b\leq -n$}

Para la soluci\'{o}n general nuevamente proponemos la combinaci\'{o}n

\begin{equation}\
w(z)=A y_{2}(a,b,z)+B y_{3}(a,b,z),
\end{equation}
ya que wronskiano de estas soluciones es

\begin{equation}\
W(y_{2},y_{3})=-\frac{\Gamma(2-b)}{\Gamma(1+a-b)}z^{-b}e^{z}
\end{equation}
Si $z\rightarrow \infty$, entonces $y_{2}\sim e^{z}z^{a-b}$ y, por
lo tanto, las autofunciones que se obtienen a partir de esta
soluci\'{o}n divergen exponencialmente.

Si $z\rightarrow 0,\quad y_{2}\sim 1$ y entonces las autofunciones
divergen como $\rho^{k}$ con $k\leq -1$ en el origen y no son de
cuadrado integrable.
\end{enumerate}

\newpage

\section*{Ap\'{e}ndice B: Campo libre en $2+1$ dimensiones}

La energ\'{\i}a de vac\'{\i}o del campo libre se obtiene de integrar las energ\'{\i}as
de la part\'{\i}cula libre en el espacio de impulsos, multiplicadas por la densidad de
estados de la part\'{\i}cula libre. Esto es

\be
  \varepsilon_{0}^{l}=\frac{E_{0}}{A}=
  -2\cdot2\frac{1}{2}\frac{1}{(2\pi)^{2}}\int
  d^{2}p\left(m^{2}+p^{2}\right)^{\frac{1}{2}}.\ee

Emplearemos el m\'{e}todo de la funci\'{o}n zeta para regularizar esta
integral, el mismo que utilizamos en la secci\'{o}n \ref{vacio-2+1}.
El par\'{a}metro $\mu$ con unidades de masa asegura que la expresi\'{o}n
tenga unidades de densidad de energ\'{\i}a por unidad de \'{a}rea.

\be
  \varepsilon_{0}^{l} =-\frac{2}{(2\pi)^{2}}\left[\mu^{s+1}
  \int_{-\infty}^{\infty}dp_{x}\int_{-\infty}^{\infty}dp_{y}
  (m^{2}+p_{x}^{2}+p_{y}^{2})^{-s/2}\right]_{s=-1}.
\end{equation}

Usamos la transformada de Mellin (\ref{mellin})

\begin{equation}
  \varepsilon_{0}^{l}=-\frac{2}{(2\pi)^{2}}\left[\frac{\mu^{s+1}}{\Gamma(s/2)}
  \int_{0}^{\infty}dt t^{s/2-1}e^{-m^{2}t}\int_{-\infty}^{\infty}dp_{x}
  e^{-p_{x}^{2}t}\int_{-\infty}^{\infty}dp_{y}e^{-p_{y}^{2}t}\right]_{s=-1},
\end{equation}
y evaluando las integrales de los impulsos con (\ref{int})

\bea
  \varepsilon_{0}^{l}&=&-\frac{2}{(2\pi)^{2}}
  \left[\frac{\mu^{s+1}}{\Gamma(s/2)}\pi
  \int_{0}^{\infty}dt t^{s/2-1-1}e^{-m^{2}t}\right]_{s=-1}\nn\\
  &&\nn\\ &=&-\frac{1}{(2\pi)}
  \left[\mu^{s+1}\frac{\Gamma(s/2-1)}{\Gamma(s/2)}(m^{2})^{-(s/2-1)}\right]_{s=-1}.
\eea

Esta expresi\'{o}n para la densidad de energ\'{\i}a de vac\'{\i}o est\'{a} regularizada
ya que la funci\'{o}n gamma s\'{o}lo presenta singularidades en los enteros negativos y
cero.

\begin{equation}
\varepsilon_{0}^{l}=-\frac{1}{(2\pi)}\frac{\Gamma\left(-3/2\right)}{\Gamma\left(-1/2\right)}m^{3}
=\frac{m^{3}}{3\pi},
\end{equation}
y el resultado coincide con el t\'{e}rmino independiente del campo del desarrollo
(\ref{vac2des}).

\newpage

\section*{Ap\'{e}ndice C: Campo libre en $3+1$ dimensiones}

De igual modo, expresamos la energ\'{\i}a de vac\'{\i}o del problema de campo libre en
$3+1$ dimensiones,

\be
  \varepsilon_{0}^{l}=\frac{E_{0}}{L_{x}L_{y}L_{z}}=
  -2\cdot2\frac{1}{2}\frac{1}{(2\pi)^{3}}\int
  d^{3}p\left(m^{2}+p^{2}\right)^{\frac{1}{2}}.\ee

Esta expresi\'{o}n es divergente. La regularizamos nuevamente por
medio del m\'{e}todo de la funci\'{o}n zeta, e introducimos un par\'{a}metro
para que conserve las unidades apropiadas

\be
  \varepsilon_{0}^{l}=-\frac{2}{(2\pi)^{3}}\left[\mu^{s+1}
  \int_{-\infty}^{\infty}dp_{x}\int_{-\infty}^{\infty}dp_{y}\int_{-\infty}^{\infty}dp_{z}
  (m^{2}+p_{x}^{2}+p_{y}^{2}+p_{z}^{2})^{-s/2}\right]_{s=-1}.
\end{equation}

Hacemos uso de la transformada de Mellin (\ref{mellin}) y obtenemos

\bea
  \varepsilon_{0}^{l}&=&-\frac{2}{(2\pi)^{3}}\left[\frac{\mu^{s+1}}{\Gamma(s/2)}
  \int_{0}^{\infty}dt t^{s/2-1}e^{-m^{2}t}\right.\times{}\nn\\ &&\nn\\
  &&\left.{}\times\int_{-\infty}^{\infty}dp_{x}
  e^{-p_{x}^{2}t}\int_{-\infty}^{\infty}dp_{y}e^{-p_{y}^{2}t}
  \int_{-\infty}^{\infty}dp_{z}e^{-p_{z}^{2}t}\right]_{s=-1}.
\eea

Resolvemos las integrales con (\ref{int}) y resulta

\bea
  \varepsilon_{0}^{l}&=&-\frac{2}{(2\pi)^{3}}
  \left[\frac{\mu^{s+1}}{\Gamma(s/2)}\pi^{3/2}
  \int_{0}^{\infty}dt t^{\frac{s-3}{2}-1}e^{-m^{2}t}\right]_{s=-1}\nn\\ &&\nn\\
  &=&-\frac{1}{4\pi^{3/2}}
  \left[\mu^{s+1}\frac{\Gamma(\frac{s-3}{2})}{\Gamma(s/2)}(m^{2})^{-\frac{s-3}{2}}\right]_{s=-1}.
\eea

Nuevamente usamos el par\'{a}metro adimensional $c$ definido por (\ref{paradi}), y
factorizamos la singularidad que la funci\'{o}n $\Gamma(\frac{s-3}{2})$ tiene en
$s=-1$,

\be
\varepsilon_{0}^{l}= -\frac{2\mu^{4}}{\pi^{3/2}}
  \left[\frac{\Gamma(\frac{s+3}{2})}{\Gamma(s/2)(s-3)(s-1)(s+1)}c^{-\frac{s-3}{2}}\right]_{s=-1},
\end{equation}
y separamos en partes singular y finita haciendo un desarrollo alrededor de la
singularidad

\be
\varepsilon_{0}^{l}=-\frac{2\mu^{4}}{\pi^{3/2}}
\left[\frac{Z(-1)}{s+1}+Z'(-1)\right]_{s=-1}, \ee donde la funci\'{o}n $Z(s)$ queda
definida por

\be
Z(s)=\frac{\Gamma(\frac{s+3}{2})c^{-\frac{s-3}{2}}}{\Gamma(s/2)(s-3)(s-1)}. \ee

Para la parte singular

\be
Z(-1)=\frac{\Gamma(1)c^{2}}{\Gamma(-1/2)(-4)(-2)}=-\frac{c^{2}}{16\sqrt{\pi}} \ee

Para el t\'{e}rmino finito

\be
Z'(-1)=-\frac{c^{2}}{32\sqrt{\pi}}\left[\psi(1)-\psi(-1/2)+\frac{3}{2}- \ln c \right].
\ee

Finalmente, para la densidad de energ\'{\i}a de vac\'{\i}o del campo libre
en $3+1$ dimensiones obtenemos

\be
\varepsilon_{0}^{l}=\frac{m^{4}}{8\pi^{2}(s+1)}+
\frac{m^{4}}{16\pi^{2}}\left[\psi(1)-\psi(-1/2)+\frac{3}{2}- \ln
\frac{m^{2}}{\mu^{2}} \right] ,\ee que coincide con el resultado
(\ref{E0libre}) obtenido en la secci\'{o}n V.

%\bibliography{biball}

\end{document}